\documentclass[conference]{IEEEtran}
\ifCLASSINFOpdf
\else
\fi
\usepackage[binary-units]{siunitx}
\DeclareSIUnit{\billion}{\text{billion}}
\usepackage[hidelinks, colorlinks=false]{hyperref}

\usepackage{tikz}
\usetikzlibrary{arrows, calc, positioning, decorations.markings, decorations.pathreplacing}
\usepackage{nicefrac}

\usepackage{pgfplots}                                                           
\usepackage{pgf}                                                           
\usepgfplotslibrary{statistics}
\usepackage{csvsimple}

\usepackage{stfloats}

\usepackage{adjustbox}
\pgfplotsset{compat=1.13}

\usepackage{amssymb} 
\usepackage{amsbsy} 

\usepackage{placeins} 

\usepackage{float}

\usepackage{xspace}
\usepackage{xcolor}
\usepackage{xifthen}
\newcommand{\ifequals}[3]{\ifthenelse{\equal{#1}{#2}}{#3}{}}

\usepackage{balance}

\newtoggle{blinded}
\togglefalse{blinded}

\iftoggle{blinded}{
\newcommand{\cvec}{\emph{CVE-redacted-3}\xspace} 
\newcommand{\cvef}{\emph{CVE-redacted-4}\xspace} 
\newcommand{\cveg}{\emph{CVE-redacted-2}\xspace} 
\newcommand{\cveh}{\emph{CVE-redacted-1}\xspace} 
\newcommand{\cvei}{\emph{CVE-redacted-5}\xspace} 
\newcommand{\cvesia}{\emph{CVE-redacted-6}\xspace} 
\newcommand{\cvesib}{\emph{CVE-redacted-7}\xspace} 
\newcommand{\cvesic}{\emph{CVE-redacted-8}\xspace} 
\newcommand{\cvesid}{\emph{CVE-redacted-9}\xspace} 
}
{
\newcommand{\cvec}{{CVE-2019-15063}\xspace} 
\newcommand{\cvef}{{CVE-2020-10370}\xspace} 
\newcommand{\cveg}{{CVE-2020-10367}\xspace} 
\newcommand{\cveh}{{CVE-2020-10368}\xspace} 
\newcommand{\cvei}{{CVE-2020-10369}\xspace} 
\newcommand{\cvesia}{{CVE-2020-29531}\xspace} 
\newcommand{\cvesib}{{CVE-2020-29533}\xspace} 
\newcommand{\cvesic}{{CVE-2020-29532}\xspace} 
\newcommand{\cvesid}{{CVE-2020-29530}\xspace} 
}

\usepackage{subcaption}
\usepackage{multirow}
\usepackage{array}
\usepackage{booktabs}

\usepackage{listings} 
\lstdefinelanguage{ASM}{
    morekeywords={b, ble, blt, bne, bx, bl, ldr, str, push, pop, mov, add, sub},
    keywordstyle=\color{blue},
    sensitive=false, 
    morecomment=[l]{//}, 
    morecomment=[s]{/*}{*/}, 
    morestring=[b]", 
} %
\lstdefinelanguage{none}{
  identifierstyle=
}

\colorlet{punct}{red!60!black}
\definecolor{background}{HTML}{EEEEEE}
\definecolor{delim}{RGB}{20,105,176}
\colorlet{numb}{magenta!60!black}

\lstdefinelanguage{json}{
    showstringspaces=false,
    breaklines=true,
    frame=lines,
    literate=
     *{:}{{{\color{punct}{:}}}}{1}
      {,}{{{\color{punct}{,}}}}{1}
      {\{}{{{\color{delim}{\{}}}}{1}
      {\}}{{{\color{delim}{\}}}}}{1}
      {[}{{{\color{delim}{[}}}}{1}
      {]}{{{\color{delim}{]}}}}{1},
}

\usepackage[nolist]{acronym}
\acrodefplural{PoC}[PoCs]{Proofs of Concept}
\acrodefplural{OS}[OSes]{Operating Systems}
\begin{acronym}
\acro{SDR}{Software-Defined Radio}
\acro{AGC}{Automatic Gain Control}
\acro{GCI}{Global Coexistence Interface}
\acro{ECI}{Enhanced Coexistence Interface}
\acro{AFH}{Adaptive Frequency Hopping}
\acro{FEM}{Front-End Module}
\acro{SP3T}{Single Pole, Triple Throw}
\acro{HCI}{Host Controller Interface}
\acro{A2DP}{Advanced Audio Distribution Profile}
\acro{SCO}{Synchronous Connection-Oriented}
\acro{GPIO}{General Purpose Input Output}
\acro{ASLR}{Address Space Layout Randomization}
\acro{LMP}{Link Management Protocol}
\acro{LCP}{Link Control Protocol}
\acro{LNA}{Low-Noise Amplifier}
\acro{EIR}{Extended Inquiry Response}
\acro{CRC}{Cyclic Redundancy Check}
\acro{RTOS}{Real-Time Operating System}
\acro{QEMU}{Quick Emulator}
\acro{UART}{Universal Asynchronous Receiver Transmitter}
\acro{MITM}{Machine-in-the-Middle}
\acro{MAC}{Media Access Control}
\acro{BLE}{Bluetooth Low Energy}
\acro{ACL}{Asynchronous Connection-Less}
\acro{DSP}{Digital Signal Processing}
\acro{SDIO}{Secure Digital Input Output}

\acro{BCS}{Bluetooth Core Scheduler}
\acro{ELF}{Executable and Linking Format}
\acro{GIAC}{Global Inquiry Access Code}
\acro{JSON}{JavaScript Object Notation}
\acro{L2CAP}{Logical Link Control and Adaptation Protocol}
\acro{LMP}{Link Management Protocol}
\acro{PTM}{Pseudo Terminal Master}
\acro{PTS}{Pseudo Terminal Slave}
\acro{QEMU}{Quick Emulator}
\acro{RCE}{Remote Code Execution}
\acro{RFU}{Reserved for Future Use}
\acro{SVC}{Supervisor Call}
\acro{UART}{Universal Asynchronous Receiver Transmitter}
\acro{WICED}{Wireless Internet Connectivity for Embedded Devices}
\acro{MMIO}{Memory Mapped Input/Output}
\acro{LE}{Bluetooth Low Energy}
\acro{IoT}{Internet of Things}
\acro{IDE}{Integrated Development Environment}
\acro{LM}{Link Manager}

\acro{RTOS}{Real-Time Operating System}
\acro{DMA}{Direct Memory Access}
\acro{RXDMA}{Receive Direct Memory Access}
\acro{NVRAM}{Non-Volatile Random-Access Memory}
\acro{ROM}{Read-Only Memory}
\acro{Rx}{Receive}
\acro{Tx}{Transmit}
\acro{SPI}{Serial Peripheral Interface}
\acro{MSP}{Main Stack Pointer}
\acro{PSP}{Process Stack Pointer}
\acro{RF}{Radio Frequency}
\acro{BLOB}{Binary Large Object}
\acro{SCO}{Synchronous Connection Oriented}
\acro{UAF}{Use-After-Free}
\acro{FHS}{Frequency Hop Sync}
\acro{CRC}{Cyclic Redundancy Check}
\acro{PDU}{Protocol Data Unit}
\acro{NFC}{Near Field Communication}
\acro{JPEG}{Joint Photographic Experts Group}
\acro{LPE}{Local Privilege Escalation}
\acro{DEP}{Data Execution Prevention}
\acro{XN}{eXecute Never}
\acro{LCP}{Link Control Protocol}
\acro{GATT}{Generic Attribute}
\acro{PoC}{Proof of Concept}
\acro{EDR}{Enhanced Data Rate}
\acro{MWS}{Mobile Wireless Standards}
\acro{HAL}{Hardware Abstraction Layer}
\acro{LR}{Link Register}
\acro{eSIM}{embedded-SIM}
\acro{RSP}{Remote SIM Provisioning}
\acro{PC}{Program Counter}
\acro{MD}{More Data}
\acro{LLID}{Logical Link Identifier}
\acro{SFI}{Serial Flash Interface}
\acro{EEPROM}{Electrically Erasable Programmable Read-Only Memory}
\acro{TOFU}{Trust On First Use}
\acro{CVE}{Common Vulnerabilities and Exposure}
\acro{PCIe}{Peripheral Component Interconnect Express}
\acro{SECI}{Serial Enhanced Coexistence Interface}
\acro{HID}{Human Interface Device}
\acro{DoS}{Denial of Service}
\acro{PTA}{Packet Traffic Arbitration}
\acro{AWMA}{Alternating Wireless Medium Access}
\acro{AP}{Access Point} 
\acro{BB}{Base Band}
\acro{SoC}{Systems on a Chip}
\acro{AWDL}{Apple Wireless Direct Link}
\acro{GAEN}{Goo\-gle Apple Exposure Notification}
\acro{SIP}{System Integrity Protection}
\acro{IOMMU}{Input--Output Memory Management Unit}
\acro{EM}{electromagnetic}
\acro{OS}{Operating System}

\end{acronym}

\usepackage[colorinlistoftodos,prependcaption,disable]{todonotes} 
\presetkeys%
    {todonotes}%
    {inline,backgroundcolor=orange}{}

\definecolor{darkred}{RGB}{212,0,16}
\definecolor{darkgreen}{RGB}{46,180,38}
\definecolor{darkergreen}{RGB}{6,98,1}
\definecolor{darkblue}{RGB}{0,73,218}

\makeatletter
\newcommand{\linebreakand}{%
  \end{@IEEEauthorhalign}
  \hfill\mbox{}\par
  \mbox{}\hfill\begin{@IEEEauthorhalign}
}
\makeatother

\begin{document}
%
\title{Attacks on Wireless Coexistence: Exploiting Cross-Technology Performance Features for Inter-Chip Privilege Escalation}

%
%
%

\author{\IEEEauthorblockN{Jiska Classen\IEEEauthorrefmark{1}}
\IEEEauthorblockA{jclassen@seemoo.de}
\and
\IEEEauthorblockN{Francesco Gringoli\IEEEauthorrefmark{2}}
\IEEEauthorblockA{francesco.gringoli@unibs.it}
\and
\IEEEauthorblockN{Michael Hermann\IEEEauthorrefmark{1}}
\IEEEauthorblockA{mhermann@seemoo.de}
\and
\IEEEauthorblockN{Matthias Hollick\IEEEauthorrefmark{1}}
\IEEEauthorblockA{mhollick@seemoo.de}
\linebreakand
\IEEEauthorblockA{\IEEEauthorrefmark{1} Technical University of Darmstadt, Secure Mobile Networking Lab\\
\IEEEauthorrefmark{2} University of Brescia, CNIT}}



%


\maketitle

%

\begin{abstract}

Modern mobile devices feature multiple wireless technologies, such as Bluetooth, Wi-Fi, and LTE.
Each of them is implemented within a separate wireless chip, sometimes packaged as combo chips. However, these chips share
components and resources, such as the same antenna or wireless spectrum.
Wireless coexistence interfaces enable them to schedule packets without collisions despite shared resources,
essential to maximizing networking performance.
Today's hardwired coexistence interfaces hinder clear security boundaries and separation between chips and chip components.
This paper shows practical coexistence attacks on \emph{Broadcom},
\emph{Cypress}, and \emph{Silicon Labs} chips deployed in billions of devices.
For example, we demonstrate that a Bluetooth chip can directly extract network passwords and manipulate traffic on a Wi-Fi chip.
Coexistence attacks enable a novel type of lateral privilege escalation across chip boundaries.
We responsibly disclosed the vulnerabilities to the vendors. Yet, only partial fixes were released for existing hardware
since wireless chips would need to be redesigned from the ground up to prevent the presented attacks on coexistence.

\end{abstract}



%


\section{Introduction}

Wireless communication is enabled by \ac{SoC}, implementing technologies such as Wi-Fi, Bluetooth, LTE, and 5G. 
While \acp{SoC} are constantly optimized towards energy efficiency, high throughput, and low latency communication, their security has not always been prioritized.
New exploits are
published continuously~\cite{quarkslab2019, 2017:artenstein, marcobh21, grant, huaweibh21, basesafe, frankenstein}.
Firmware patching to mitigate flaws requires strong collaboration between vendors and manufacturers, leading
to asynchronous, incomplete, and slow patch cycles~\cite{frankenstein}. In addition, firmware patch diffing
can provide attackers with \ac{SoC} vulnerabilities multiple months before public disclosure~\cite{polypyus}.

Mobile device vendors account for potentially insecure wireless \acp{SoC} by isolating them
from the \ac{OS} and hardening the \ac{OS} against escalation strategies.
For example, on \emph{Android}, the Bluetooth daemon residing on top of \ac{OS} drivers 
runs with limited privileges, is sandboxed, and is currently being reimplemented in a memory-safe language~\cite{gabeldorsche}.
As a result, recent wireless exploit chains targeting the mobile \ac{OS} instead of the \ac{SoC} are rather complex
and need to find a bypass for each mitigation~\cite{2020:googleprojectzero, 2017:googleprojectzero, bluefrag}.

We provide empirical evidence that coexistence, i.e., the coordination of cross-technology wireless transmissions, is an unexplored attack surface.
Instead of escalating directly into the mobile \ac{OS}, wireless chips can escalate their privileges into other wireless chips by exploiting the same mechanisms they use to arbitrate their access to the resources they share, i.e., the transmitting antenna and the wireless medium. This new model of wireless system exploitation is comparable to well-known threats that occur when multiple threads or users can share resources like processors or memory~\cite{rowhammer, spectre, meltdown}.
This paper demonstrates lateral privilege escalations from a Bluetooth chip to code execution on a Wi-Fi chip.
The Wi-Fi chip encrypts network
traffic and holds the current Wi-Fi credentials, thereby providing the attacker with further information.
Moreover, an attacker can execute code on a \mbox{Wi-Fi} chip even if it is not connected to a wireless network.
In the opposite direction, we observe Bluetooth packet types from a Wi-Fi chip. This allows
determining keystroke timings on Bluetooth keyboards, which can allow reconstructing texts
entered on the keyboard~\cite{brokenstrokes}.

Since wireless chips communicate directly through hardwired coexistence interfaces, the \textbf{\ac{OS} drivers cannot filter any events to
prevent this novel attack}. Despite reporting the first security issues on these
interfaces more than two years ago, the inter-chip interfaces remain vulnerable to most
of our attacks. For instance, Bluetooth$\rightarrow$\mbox{Wi-Fi} code execution is still possible on \emph{iOS 14.7} and \emph{Android 11}.

Wireless coexistence is indispensable for high-performance wireless transmissions on any modern device~\cite{coexperf}.
We experimentally confirm that these interfaces exist and are vulnerable: 
\begin{itemize}
\item We explore a \emph{Broadcom} and
      \emph{Cypress} Bluetooth$\leftrightarrow$\mbox{Wi-Fi} interface, which is present in all \emph{iPhones}
      and \emph{MacBooks}, the \emph{Samsung Galaxy S} series, \emph{Raspberry Pis} and IoT devices.
\item We successfully launch Bluetooth$\rightarrow$Wi-Fi code execution on all recent
	  \emph{Broadcom} and \emph{Cypress} combo chips. 
\item We implement packet type information disclosure and \ac{DoS} on the standardized IEEE 802.15.2
      coexistence interface, used by \emph{Silicon Labs} and further vendors, to show that this novel attack type is generally applicable.
\end{itemize}
We received nine \acf{CVE} identifiers for the implemented attacks (see \autoref{tab:overview}).
Since the underlying issue is broader than our practical attack implementations, we informed the
\emph{Bluetooth SIG}, responsible for the \ac{MWS} specification. Moreover, we included wireless chip manufacturers, such as 
\emph{Intel}, \emph{Marvell}, \emph{MediaTek}, \emph{NXP}, \emph{Qualcomm}, and \emph{Texas Instruments},
in the responsible disclosure process, if their datasheets mentioned coexistence interfaces~\cite{coexintel,coexmediatek,coexmarvell,coexqualcomm,coexti}.

The rest of the paper is structured as follows.
We provide a background on wireless coexistence in \autoref{sec:coexdef}.
Based on the threat model presented in \autoref{sec:threatmodel}, we introduce coexistence attack concepts
in \autoref{sec:attackconcepts}.
Then, we detail practical coexistence attacks on various chips and technologies in \autoref{sec:practial}.
In \autoref{sec:mitigation}, we discuss patch timelines
and mitigation strategies while hardware-based fixes are not available.
We discuss other wireless side channels, which are related to our novel coexistence-based attack scheme, in \autoref{sec:relwork}.
We conclude our work in \autoref{sec:conclusion}.


\section{Wireless Coexistence}
\label{sec:coexdef}

Minimizing the number of collisions is a fundamental goal of all wireless communication systems. Technologies like Bluetooth and LTE prevent collisions by assigning to each node a fixed schedule of reserved time slots~\cite{bt52}. Wi-Fi nodes, instead, try to avoid collisions by repeatedly deferring their transmissions if the channel is busy~\cite{bianchi}. While these mechanisms are effective when working in isolation, different channel access mechanisms operating in close proximity over the same frequencies can cause severe interference~\cite{spork2020improving}. Wireless coexistence solves this problem by introducing an additional control layer that arbitrates transmissions from different transceivers collocated in the same node~\cite{coexperf}.

\tikzset{>=latex}
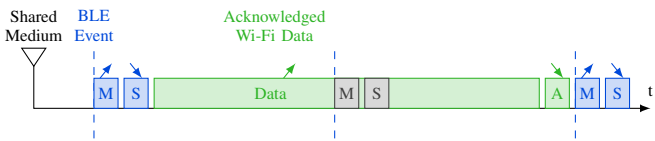
\begin{figure}[bp]
\vspace{-1em} 
	\center
	\begin{tikzpicture}[minimum height=0.55cm, scale=0.8, every node/.style={scale=0.8}, node distance=0.7cm]

    \draw[<-] (11.25,0.5) -- (1,0.5) -- (1, 1.2) -- (0.8, 1.5) -- (1.2, 1.5) -- (1, 1.2);
    \node[align=center,anchor=south,font=\footnotesize] (txtsa) at (1, 1.5) {Shared \\ Medium};
    \node[align=center,anchor=south,font=\footnotesize] (txtt) at (11.25, 0.5) {t};
    
    \draw[-,dashed,darkblue] (2,0) -- (2,1.5);
    \node[align=center,anchor=south,font=\footnotesize,darkblue] (txtbs) at (2, 1.5) {BLE \\ Event};
    \draw[-,dashed,darkblue] (6,0) -- (6,1.5);
    \draw[-,dashed,darkblue] (10,0) -- (10,1.5);
    
    \filldraw[fill=darkblue!20,draw=darkblue](2,0.5) rectangle node (bt) {\footnotesize{\textcolor{darkblue}M}} ++(0.4,0.5);
    \draw[->,darkblue](2.1,1) -- (2.3,1.25);
    \filldraw[fill=darkblue!20,draw=darkblue](2.5,0.5) rectangle node (bt) {\footnotesize{\textcolor{darkblue}S}} ++(0.4,0.5);
    \draw[<-,darkblue](2.8,1) -- (2.6,1.25);
    
    \filldraw[fill=darkblue!20,draw=darkblue](10,0.5) rectangle node (bt) {\footnotesize{\textcolor{darkblue}M}} ++(0.4,0.5);
    \draw[->,darkblue](10.1,1) -- (10.3,1.25);
    \filldraw[fill=darkblue!20,draw=darkblue](10.5,0.5) rectangle node (bt) {\footnotesize{\textcolor{darkblue}S}} ++(0.4,0.5);
    \draw[<-,darkblue](10.8,1) -- (10.6,1.25);

    \node[align=center,anchor=south,font=\footnotesize,darkgreen] (txtbs) at (5, 1.5) {Acknowledged \\ Wi-Fi Data};
    
    \filldraw[fill=darkgreen!20,draw=darkgreen](3,0.5) rectangle node (bt) {\footnotesize{\textcolor{darkgreen}{Data\hspace*{9em}}}} ++(6.4,0.5);
    \draw[->,darkgreen](5.15,1) -- (5.35,1.25);
    \filldraw[fill=darkgreen!20,draw=darkgreen](9.5,0.5) rectangle node (bt) {\footnotesize{\textcolor{darkgreen}A}} ++(0.4,0.5);
    \draw[<-,darkgreen](9.8,1) -- (9.6,1.25);

    \filldraw[fill=darkgray!20,draw=darkgray](6,0.5) rectangle node (bt) {\footnotesize{\textcolor{darkgray}M}} ++(0.4,0.5);
    \filldraw[fill=darkgray!20,draw=darkgray](6.5,0.5) rectangle node (bt) {\footnotesize{\textcolor{darkgray}S}} ++(0.4,0.5);

	\end{tikzpicture}
\vspace{-0.5em} 
\caption{Wireless packet scheduling example for Wi-Fi and BLE on a shared medium.}
\label{fig:coexpackets}
\end{figure}

\subsection{Working Principle}

We show in \autoref{fig:coexpackets} \ac{BLE} (blue) and Wi-Fi (green) frames transmitted and received at a node that adopts a wireless coexistence mechanism. When BLE is operating using in its assigned time slots, the Wi-Fi transceiver is literally blocked. As soon as the BLE transceiver releases the channel, the Wi-Fi chip may start transmitting a data frame and afterward wait for the acknowledgment that it expects from the receiver. Meanwhile, the BLE transceiver cannot use any time slot that may occur (shadowed gray frames in the middle of the Wi-Fi one) and rests until the next Wi-Fi-free slot.
Such coordination is fundamental to \emph{i)} prevent collisions that neither BLE nor Wi-Fi would have been able to avoid on their own and hence \emph{ii)} increase the aggregated throughput. 

Also, technologies that use different frequencies can collide and must adopt some coexistence mechanism. For example, 3GPP allocated for the uplink channel of LTE band 7 frequencies above \SI{2.5}{\giga\hertz} that are very close to the upper side of the \SI{2.4}{\giga\hertz} ISM band used by Wi-Fi and Bluetooth: interference is still possible because of spurious harmonics even if there is no actual spectral overlapping~\cite{coex_appnote}.

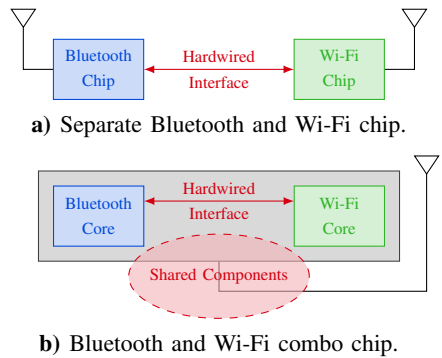
\begin{figure}[bp]
\vspace{-1em} 
\centering
\begin{subfigure}{0.5\textwidth}
	\center
	\begin{tikzpicture}[minimum height=0.55cm, scale=0.8, every node/.style={scale=0.8}, node distance=0.7cm] 
	
	  \begin{scope}[shift={(4.8,0)}] 
	    \draw[-] (-0.3,0.5) -- (0.2,0.5) -- (0.2, 1.2) -- (0, 1.5) -- (0.4, 1.5) -- (0.2, 1.2);
	  \end{scope}
	  \begin{scope}[shift={(-1.7,0)}] 
	    \draw[-] (0.7,0.5) -- (0.2,0.5) -- (0.2, 1.2) -- (0, 1.5) -- (0.4, 1.5) -- (0.2, 1.2);
	  \end{scope}
	
    \filldraw[fill=darkblue!20,draw=darkblue](-1,0) rectangle node[align=center] (bt) {\footnotesize{\textcolor{darkblue}{Bluetooth}} \\ \footnotesize{\textcolor{darkblue}{Chip}}} ++(1.5,1);
    \filldraw[fill=darkgreen!20,draw=darkgreen](3,0) rectangle node[align=center] (bt) {\footnotesize{\textcolor{darkgreen}{Wi-Fi}} \\ \footnotesize{\textcolor{darkgreen}{Chip}}} ++(1.5,1);
    \path[<->,color=darkred] (0.5,0.5) edge node[sloped,yshift=0cm,align=center] {\footnotesize{Hardwired}\\ \footnotesize{Interface}} (3,0.5);

	\end{tikzpicture}

\caption{Separate Bluetooth and Wi-Fi chip.}
\label{fig:coexbtwifisep}

\end{subfigure}

\begin{subfigure}{0.5\textwidth}
	\center
	\hspace{0.4cm} 
	\begin{tikzpicture}[minimum height=0.55cm, scale=0.8, every node/.style={scale=0.8}, node distance=0.7cm]

	  \begin{scope}[shift={(5,0)}] 
	    \draw[-] (-3.25,0) -- (-3.25,-0.75) -- (0.2,-0.75) -- (0.2, 1.2) -- (0, 1.5) -- (0.4, 1.5) -- (0.2, 1.2);
	  \end{scope}
	  
	\filldraw[fill=darkgray!20,draw=darkgray](-1.25,-0.25) rectangle node[align=center] (bt) {} ++(6,1.5);
	
    \filldraw[fill=darkblue!20,draw=darkblue](-1,0) rectangle node[align=center] (bt) {\footnotesize{\textcolor{darkblue}{Bluetooth}} \\ \footnotesize{\textcolor{darkblue}{Core}}} ++(1.5,1);
    \filldraw[fill=darkgreen!20,draw=darkgreen](3,0) rectangle node[align=center] (bt) {\footnotesize{\textcolor{darkgreen}{Wi-Fi}} \\ \footnotesize{\textcolor{darkgreen}{Core}}} ++(1.5,1);
    \path[<->,color=darkred] (0.5,0.75) edge node[sloped,yshift=0cm,align=center] {\footnotesize{Hardwired}\\ \footnotesize{Interface}} (3,0.75);
    
    \draw[color=darkred, dashed, fill=darkred!30, fill opacity=0.6] (1.75,-0.5) ellipse (1.5cm and 0.7cm) node[align=center, fill opacity=1] (cx) {\footnotesize{Shared Components}};

	\end{tikzpicture}

\caption{Bluetooth and Wi-Fi combo chip.}
\label{fig:coexbtwificomb}
\end{subfigure}
\vspace{-0.5em} 
\caption{Coexistence architectures.}
\label{fig:coexbtwifiarch}
\end{figure}

\subsection{Coexistence Architecture}

Coexistence implementations depend on the underlying chip architecture:
separate chips or combo chips.
On separate chips (see \autoref{fig:coexbtwifisep}), all scheduling information
is exchanged using an external hardwired interface. Sometimes, such interfaces
are documented for interoperability between chips by different vendors.
Combo chips (see \autoref{fig:coexbtwificomb}) typically embed separate cores dedicated to the different protocols for performance reasons. The combined
design allows sharing redundant components such as the antenna. However, 
additional shared components, such as transmission and reception buffers, could also
be involved in coexistence coordination. It is up to the vendor to optimize the hardware
design and add proprietary features.

Both concepts are common in practice. For example, we reverse-engineered coexistence
interfaces on an \emph{iPhone 11}. We found that Bluetooth and \mbox{Wi-Fi} use a
\emph{Broadcom} combo chip with proprietary coexistence features, while the \emph{Intel} LTE
baseband chip is interconnected with a serial \ac{MWS} interface.

\subsection{Involved Components}

Coexistence interfaces are part of the chip's hardware because
packet scheduling requires real-time information from multiple chips or cores.
The firmware running on the chip can send data to hardwired interfaces by writing to \ac{MMIO}
addresses. Thus, some coexistence behavior can be firmware-defined.
Configurability depends on the initial hardware design by the vendor.

In general, coexistence is managed directly by the chips without involving the \ac{OS}, e.g. \emph{Android} or \emph{iOS}, which does not see real-time packet scheduling information. However, coexistence information that is not timing-critical can be coordinated externally. For example, on an \emph{iPhone} with \ac{MWS}, frequency band configuration information from the LTE chip is processed by \emph{iOS} and then forwarded to the Bluetooth chip.

\subsection{Relevance}

Due to backward compatibility, coexistence features are bound to be available on future chips,
even if chunks of Wi-Fi traffic are offloaded to the \SI{5}{\giga\hertz} or \SI{60}{\giga\hertz} bands.
Even now, recent standards such as 802.11ax continue to operate in the \SI{2.4}{\giga\hertz} band.

\begin{table*}[tp]
\renewcommand{\arraystretch}{1.3}
\caption{Practical coexistence attacks demonstrated in this paper.}
\label{tab:overview}
\vspace{-0.5em} 
\centering
\scriptsize
\begin{tabular}{|l|l|l|l|l|l|c|}
\hline
\textbf{Attack Type} & \textbf{Component} & \textbf{Chip Type} & \textbf{Vendor} & \textbf{CVE} & \textbf{Impact} & \textbf{Patchable}\\
\hline
\multirow{2}{*}{Architectural} & \multirow{2}{*}{Shared Memory} & \multirow{2}{*}{Combo} & \multirow{2}{*}{Broadcom, Cypress} & {\cveh} & Bluetooth$\rightarrow$Wi-Fi memory readout & $\pmb{\times}$\\
& & & & {\cveg} & Bluetooth$\rightarrow$Wi-Fi code execution & $\pmb{\times}$\\
\hline
\multirow{7}{*}{Protocol-based} & \multirow{3}{*}{SECI (Proprietary)} & \multirow{3}{*}{Combo, Separate} & \multirow{3}{*}{Broadcom, Cypress} & {\cvec} & Bluetooth$\rightarrow$Wi-Fi denial of service & $\pmb{\checkmark}$ \\
& & & & {\cvef} & Wi-Fi$\rightarrow$Bluetooth denial of service & $\pmb{\checkmark}$ \\
& & & & {\cvei} & Wi-Fi$\rightarrow$Bluetooth packet timing \& metadata & \textbf{?} \\ \cline{2-7}
& \multirow{4}{*}{PTA (Standardized)} & \multirow{4}{*}{Separate} & \multirow{4}{*}{Silicon Labs} & {\cvesia} & Bluetooth$\rightarrow$Wi-Fi denial of service & $\pmb{\checkmark}$ \\
& & & & {\cvesib} & Bluetooth$\rightarrow$Wi-Fi packet timing \& metadata & \textbf{?} \\
& & & & {\cvesic} & Wi-Fi$\rightarrow$Bluetooth denial of service & $\pmb{\checkmark}$ \\
& & & & {\cvesid} & Wi-Fi$\rightarrow$Bluetooth packet timing \& metadata & \textbf{?}\\
\hline
\end{tabular}
\vspace{-0.5em} 
\end{table*}

\section{Threat Model}
\label{sec:threatmodel}

In the following, we explain the attacker's goals and prerequisites for launching
coexistence attacks.

\subsection{Attacker Goals}
When exploiting coexistence, the attacker aims at escalating privileges laterally from one
wireless chip or core into another. This way, the attacker can extract secret information
only available to this other chip (e.g., steal a Wi-Fi password via Bluetooth) and generally
widens the attack surface for further escalating into the \ac{OS} (e.g., use
a Bluetooth chip vulnerability to escalate into an \emph{Android} Wi-Fi driver).

\subsection{Attacker Prerequisites}

We assume an attacker with code execution on one wireless component.
Thus, one wireless core or chip is untrusted.
While this is a strong attack precondition, it is
similar to other side-channel attacks that escalate privileges
for information extraction or code execution, e.g., executing \emph{JavaScript} on a website~\cite{rowhammerjs, spectrejs}.

Controlling a wireless component requires one of the following efforts. An attacker can execute code by \emph{i)} exploiting an unpatched
or new security issue over-the-air, or \emph{ii)} use the local \ac{OS} firmware update mechanism~\cite{mantz2019internalblue,nexmon:project}.
Coexistence behavior can also be observed on higher layers, e.g., \emph{iii)} untrusted applications on a smartphone can observe side-effects of packet scheduling of either Bluetooth or Wi-Fi. Such observations do not require code execution within a wireless chip but exploit the mere existence of coexistence protocols.
In the remainder of this paper, we consider an over-the-air attacker \emph{i)}.
We will practically confirm our attacks using the code execution methods \emph{i)} and \emph{ii)}.
We neglect \emph{iii)}, since it does not directly involve wireless chips.

Typically, when a new exploitable wireless bug or attack class is found, it only affects one wireless
technology, such as Bluetooth~\cite{braktooth,sweyntooth,frankenstein,veronica,mantz2019internalblue}, Wi-Fi~\cite{quarkslab2019,2017:artenstein,2017:googleprojectzero,owfuzz,defconqualcomm}, or cellular  basebands~\cite{amat,nicointel,marcobh21,guy2019,grant,huaweibh21,basesafe,grassi}. 
Coexistence attacks change this in favor of the attacker: vulnerabilities of a single technology can affect multiple technologies due to the lateral privilege escalation.

Even without finding a new bug, an attacker has a chance to gain initial over-the-air code execution due to patch gapping.
Wireless firmware patching requires close collaboration between \ac{OS} maintainers and
hardware manufacturers. In practice, rolling out such patches takes multiple months, and some devices receive patches earlier than others~\cite{frankenstein}.
Despite coordinated disclosure processes, even in 2021, Bluetooth chip vulnerability descriptions
have been released prior to patches~\cite{braktooth}, resulting in a patch gap of approximately 1--2 months, depending on the vendor. Similarly, a Wi-Fi fuzzing tool capable of finding new bugs
was released in 2021 
prior to rolling out all patches~\cite{owfuzz}. 
After such publications, the specific technology is at high risk of being
exploited, until the vendor rolls out patches and users install them.

\section{Coexistence Attack Concepts}
\label{sec:attackconcepts}

We implement practical coexistence attacks against popular
chips, as listed in \autoref{tab:overview}.
Prior to implementing these attacks, we provide a brief description
of their root cause and impact.
The first two vulnerabilities are \emph{architectural}, while the remaining
vulnerabilities are \emph{protocol-based}.

\subsection{Architectural Vulnerabilities}

Wireless cores or chips can share otherwise redundant components.
Even though there is no detailed documentation, some datasheets
briefly mention shared components like the antenna~\cite{coexmarvell,coexqualcomm,coexti}, the USB to host interface~\cite{coexmediatek}, or memory~\cite{bcm4339}.
Chip architectures might contain further undocumented shared components.

Within a shared component, data of multiple wireless technologies is mixed, e.g., in a shared
buffer that caches raw samples prior to transmission via the antenna, or in a shared interface to the
\ac{OS} such as USB or PCIe. If these components miss separation, one technology can
access data of the other technology. Moreover, the \ac{OS} sends data to the chips in plaintext,
and chips perform encryption, meaning that \textbf{unencrypted data can leak} over shared components (\cveh).
Additionally, if shared components hold firmware data or are involved in firmware updates,
this can lead to \textbf{code execution} (\cveg).

\subsection{Protocol-based Vulnerabilities}

Separate wireless chips use hardwired serial interfaces to coordinate packet transmissions.
Such interfaces can also be part of a combo chip design, in addition to shared components.

Even a simple serial protocol, which only allows or defers transmissions, can be
susceptible to \textbf{\ac{DoS}}, e.g., the coexistence interface can be misconfigured on one end (\cvec), transmission requests can be ignored (\cvef), or priority transmissions can be requested permanently (\cvesia, \cvesic).

Most serial coexistence protocols contain additional
information to optimize scheduling, such as transmission slot timings and packet priorities.
This can \textbf{leak information} across wireless chips: depending on the underlying serial protocol an attacker can determine current packet types and activity (\cvesib, \cvesid) or even understand if a Bluetooth packet contained keystrokes (\cvei).

\subsection{Stealthiness}
Since wireless chip code execution flaws are published so frequently, \acp{OS} harden their wireless stacks against privilege escalations from the chip towards the \ac{OS}, e.g., by reimplementing wireless daemons in \emph{Rust}~\cite{gabeldorsche}. However, instead of escalating into higher-level components of wireless stacks, our attacker aims at escalating laterally between wireless technologies within the same level (wireless chip). On this level, components are hardwired, and the \ac{OS} can neither observe nor filter an ongoing attack. 
Thus, coexistence flaws bypass \ac{OS}-level protections like sandboxing wireless daemons.

\subsection{Patchability}

Some issues can only be patched by releasing a new hardware revision.
For example, a new firmware version will not physically remove shared memory from a chip
or adjust for arbitrary jitter in a serial protocol.
Moreover, some packet timing and metadata cannot be removed without 
negatively impacting packet coordination performance. 
More details on patches and the disclosure timeline follow in \autoref{ssec:patches}.

\tikzset{>=latex}
\begin{figure*}[bp]
\vspace{-1em} 

	\begin{tikzpicture}[minimum height=0.55cm, scale=0.8, every node/.style={scale=0.8}, node distance=0.7cm]

    \filldraw[fill=darkblue!20,draw=darkblue,thick](-8.1,-3) rectangle node (bt) {} ++(5.5,7.1);
    \node[anchor=north,align=center, above=of bt.north,yshift=2.5cm] (chiptxtb) {\textbf{Bluetooth}};

    \filldraw[fill=gray!10,draw=gray!50](-7.75,1.8) rectangle node (btpcx) {\rotatebox{90}{\footnotesize{Port Control}}} ++(0.4,1.95);
    \filldraw[fill=gray!10,draw=gray!50](-7,1.8) rectangle node (btpc) {} ++(1.5,1.95);
    \filldraw[fill=white](-6.75,3.25) rectangle node (btpca) {\footnotesize{UART}} ++(1,0.3);
    \filldraw[fill=white](-6.75,2.85) rectangle node (btpcb) {\footnotesize{I2S}} ++(1,0.3);
    \filldraw[fill=white](-6.75,2.45) rectangle node (btpcc) {\footnotesize{PCM}} ++(1,0.3);
    \filldraw[fill=white](-6.75,2.05) rectangle node (btpcd) {\footnotesize{USB}} ++(1,0.3);
    \draw[<->] (-7.35,3.4)--(-6.75,3.4);
    \draw[<->] (-7.35,3)--(-6.75,3);
    \draw[<->] (-7.35,2.6)--(-6.75,2.6);
    \draw[<->] (-7.35,2.2)--(-6.75,2.2);
    
    \draw[-,gray,dashed] (-8.5,4.2) -- (-8.5,1.3);
    \node[align=right] (pco) at (-10, 2.8) {\footnotesize{BT\_HOST\_WAKE}\\ \footnotesize{BT\_DEV\_WAKE}\\ \footnotesize{UART}\\ \footnotesize{USB 2.0}\\ \footnotesize{PCM}\\ \footnotesize{I$^2$S}\\ \footnotesize{Other GPIOs}};
    \draw[<->] (-8.8,2.8)--(-7.75,2.8);
    
    \draw[-,gray,dashed] (-8.5,-2.3) -- (-8.5,-2.9);
    \node[align=right] (lpo) at (-10.2, -2.6) {\footnotesize{32\,kHz External LPO}};
    \draw[->] (-8.8,-2.6)--(-7.75,-2.6);
    
    \filldraw[fill=gray!10,draw=gray!50](-7,1.05) rectangle node (regs) {\footnotesize{Registers}} ++(1.5,0.4);
    \filldraw[fill=gray!10,draw=gray!50](-7,0.55) rectangle node (dma) {\footnotesize{DMA}} ++(1.5,0.4);
    \filldraw[fill=gray!10,draw=gray!50](-7,0.05) rectangle node (jtag) {\footnotesize{JTAG}} ++(1.5,0.4);

    \filldraw[fill=gray!10,draw=gray!50](-7.75,-2.25) rectangle node (ahb) {} ++(1.5,1.95);
    \filldraw[fill=white](-7.5,-0.8) rectangle node (ahba) {\footnotesize{GPIO}} ++(1,0.3);
    \filldraw[fill=white](-7.5,-1.2) rectangle node (ahbb) {\footnotesize{Timers}} ++(1,0.3);
    \filldraw[fill=white](-7.5,-1.6) rectangle node (ahbc) {\footnotesize{WD}} ++(1,0.3);
    \filldraw[fill=white](-7.5,-2) rectangle node (ahbd) {\footnotesize{Pause}} ++(1,0.3);
    \filldraw[fill=gray!10,draw=gray!50](-5.9,-2.25) rectangle node (ahbx) {\rotatebox{90}{\footnotesize{AHB2APB}}} ++(0.4,1.95);
    \draw[<-] (-6.5,-0.65)--(-5.9,-0.65);
    \draw[<-] (-6.5,-1.05)--(-5.9,-1.05);
    \draw[<-] (-6.5,-1.45)--(-5.9,-1.45);
    \draw[<-] (-6.5,-1.85)--(-5.9,-1.85);
    
    \filldraw[fill=gray!10,draw=gray!50](-5.15,-2.25) rectangle node (ahbm) {\rotatebox{90}{\footnotesize{AHB Bus Matrix}}} ++(0.4,6);
    \draw[<-] (-5.5,2.8)--(-5.15,2.8);
    \draw[<-] (-5.5,1.25)--(-5.15,1.25);
    \draw[->] (-5.5,0.75)--(-5.15,0.77);
    \draw[->] (-5.5,0.25)--(-5.15,0.25);
    \draw[<-] (-5.5,-1.25)--(-5.15,-1.25);
    \draw[->] (-4.75,3.4)--(-4.4,3.4);
    \draw[->] (-4.75,3)--(-4.4,3);
    \draw[<-] (-4.75,2.3)--(-4.4,2.3);
    \draw[<-] (-4.75,2.45)--(-4.4,2.45);
    \draw[<-] (-4.75,2.6)--(-4.4,2.6);
    \draw[<-] (-4.75,2.05)--(-4.4,2.05);
    \draw[->] (-4.75,1.85)--(-4.4,1.85);
    \draw[->] (-4.75,0.3)--(-4.4,0.3);

    \filldraw[fill=gray!10,draw=gray!50, align=center](-4.4,2.75) rectangle node (regs) {\footnotesize{RAM} \\ \footnotesize{ROM}} ++(1.5,1);
    \filldraw[fill=gray!10,draw=gray!50, align=center](-4.4,2.25) rectangle node (regs) {\footnotesize{ARM CM3}} ++(1.5,0.4);
    \filldraw[fill=gray!10,draw=gray!50, align=center](-4.4,1.75) rectangle node (regs) {\footnotesize{WLAN M/S}} ++(1.5,0.4);

    \begin{scope}[xshift=3.35cm,yshift=1.75cm]
    \filldraw[fill=gray!10,draw=gray!50](-7.75,-2.65) rectangle node (rxx) {} ++(1.5,2.35);
    \filldraw[fill=white](-7.5,-0.8) rectangle node (rxa) {\footnotesize{RX/TX}} ++(1,0.3);
    \filldraw[fill=white](-7.5,-1.2) rectangle node (rxb) {\footnotesize{BLE}} ++(1,0.3);
    \filldraw[fill=white](-7.5,-1.6) rectangle node (rxc) {\footnotesize{LCU}} ++(1,0.3);
    \filldraw[fill=white](-7.5,-2) rectangle node (rxd) {\footnotesize{APU}} ++(1,0.3);
    \filldraw[fill=white](-7.5,-2.4) rectangle node (rxd) {\footnotesize{BlueRF}} ++(1,0.3);
    \end{scope}
    
    \begin{scope}[xshift=3.35cm,yshift=-0.85cm]
    \filldraw[fill=gray!10,draw=gray!50](-7.75,-1.4) rectangle node (mx) {} ++(1.5,1.15);
    \filldraw[fill=white](-7.5,-0.8) rectangle node (mxa) {\footnotesize{Modem}} ++(1,0.3);
    \filldraw[fill=white](-7.5,-1.2) rectangle node (mxb) {\footnotesize{BT RF}} ++(1,0.3);
    \end{scope}
    \draw[->] (-3.7,-0.9)--(-3.7,-1.35); 
    \draw[->] (-4,-2.05)--(-4,-3.1)--(2.5,-3.1)--(2.5,-3.3); 
    
    \fill[fill=white,draw=gray!50,yshift=-2.6cm,xshift=-0.2cm] (-4,0)--(-3.6,0)--(-3.8,-0.2)--(-4,0);
    \node[align=left] (btpa) at (-3.25, -2.7) {\footnotesize{BT PA}};

    
    \filldraw[fill=darkgreen!20,draw=darkgreen,thick](2,-3) rectangle node (wifi) {} ++(4.7,7.1);
    \node[anchor=north,align=center, above=of wifi.north,yshift=2.5cm] (chiptxtw) {\textbf{Wi-Fi}};
    \draw[] (3.05,2.75)--(3.05,2.25);
    \draw[] (4.35,0.45)--(4.35,-2.2);

    \filldraw[fill=gray!10,draw=gray!50, align=center](2.3,2.75) rectangle node (regs) {\footnotesize{RAM} \\ \footnotesize{ROM}} ++(1.5,1);
    \filldraw[fill=gray!10,draw=gray!50, align=center](2.3,2.25) rectangle node (regs) {\footnotesize{ARM CR4}} ++(1.5,0.4);
    \filldraw[fill=gray!10,draw=gray!50, align=center](2.3,1.75) rectangle node (regs) {\footnotesize{AXI$\leftrightarrow$AHB}} ++(1.5,0.4);
    \filldraw[fill=gray!10,draw=gray!50, align=center](2.3,0.45) rectangle node (regs) {\footnotesize{Chip}\\ \footnotesize{Common}\\ \footnotesize{OTP}} ++(1.5,1.2);
    \filldraw[fill=gray!10,draw=gray!50, align=center](2.3,-0.45) rectangle node (regs) {\footnotesize{D11 Core (MAC)}} ++(4.1,0.8);
    \filldraw[fill=gray!10,draw=gray!50, align=center](2.3,-1.35) rectangle node (regs) {\footnotesize{1$\times$1 802.11ac PHY}} ++(4.1,0.8);
    \filldraw[fill=gray!10,draw=gray!50, align=center](2.3,-2.25) rectangle node (regs) {\footnotesize{2.4\,GHz/5\,GHz 802.11ac}\\ \footnotesize{Dual-Band Radio}} ++(4.1,0.8);
    
    \draw[->] (3.8,2.55)--(4.15,2.55); 
    \draw[<-] (3.8,2.35)--(4.15,2.35); 
    \draw[->] (3.8,2.05)--(4.15,2.05); 
    \draw[<-] (3.8,1.85)--(4.15,1.85); 
    \draw[->] (3.8,1.15)--(4.15,1.15); 
    \draw[<-] (3.8,0.95)--(4.15,0.95);

    \draw[<-] (4.55,1.65)--(4.9,1.65); 
    \draw[->] (4.55,1.25)--(4.9,1.25); 
    \draw[<-] (4.55,1.05)--(4.9,1.05); 
    \draw[->] (4.55,0.65)--(4.9,0.65); 

    \filldraw[fill=gray!10,draw=gray!50](4.15,0.45) rectangle node (nic) {\rotatebox{90}{\footnotesize{NIC-301 AXI Backplane}}} ++(0.4,3.3);
    
    \filldraw[fill=gray!10,draw=gray!50, align=center](4.9,1.45) rectangle node (regs) {\footnotesize{SDIO D}} ++(1.5,0.4);
    \filldraw[fill=gray!10,draw=gray!50, align=center](4.9,0.95) rectangle node (regs) {\footnotesize{PCIe}} ++(1.5,0.4);
    \filldraw[fill=gray!10,draw=gray!50, align=center](4.9,0.45) rectangle node (regs) {\footnotesize{AXI2APB}} ++(1.5,0.4);

    \begin{scope}[xshift=-0.1cm]
    \draw[->,yshift=-2.5cm,xshift=7.5cm] (-3.8,0.25)--(-3.8,-0.8);
    \fill[fill=white,draw=gray!50,yshift=-2.6cm,xshift=7.5cm] (-4,0)--(-3.6,0)--(-3.8,-0.2)--(-4,0);
    \draw[<-,yshift=-2.5cm,xshift=7cm] (-3.8,0.25)--(-3.8,-0.8);
    \fill[fill=white,draw=gray!50,yshift=-2.6cm,xshift=7cm] (-4,-0.2)--(-3.6,-0.2)--(-3.8,0)--(-4,-0.2);
    \draw[->,yshift=-2.5cm,xshift=10cm] (-3.8,0.25)--(-3.8,-0.8);
    \fill[fill=white,draw=gray!50,yshift=-2.6cm,xshift=10cm] (-4,0)--(-3.6,0)--(-3.8,-0.2)--(-4,0);
    \draw[<-,yshift=-2.5cm,xshift=9cm] (-3.8,0.25)--(-3.8,-0.8);
    \fill[fill=white,draw=gray!50,yshift=-2.6cm,xshift=9cm] (-4,-0.2)--(-3.6,-0.2)--(-3.8,0)--(-4,-0.2);
	\end{scope}
	
    \draw[-,gray,dashed] (7.1,4) -- (7.1,-2.25);
    \node[align=left, anchor=west] (wpo) at (7.4, 3.2) {\footnotesize{WL\_HOST\_WAKE}\\ \footnotesize{WL\_DEV\_WAKE}\\ \footnotesize{JTAG}\\ \footnotesize{Other GPIOs}};
    \draw[<->] (6.3,3.2)--(7.4,3.2);
    \node[align=left, anchor=west] (sdio) at (7.4, 1.65) {\footnotesize{SDIO 3.0}};
    \draw[<-] (6.4,1.65)--(7.4,1.65);
    \node[align=left, anchor=west] (pcie) at (7.4, 1.15) {\footnotesize{PCIe 1.1}};
    \draw[<-] (6.4,1.15)--(7.4,1.15);
    \node[align=left, anchor=west] (rfs) at (7.4, -0.95) {\footnotesize{RF Switch Controls}};
    \draw[->] (6.4,-0.95)--(7.4,-0.95);
    \node[align=left, anchor=west] (xtal) at (7.4, -1.85) {\footnotesize{XTAL}};
    \draw[<-] (6.4,-1.85)--(7.4,-1.85);
    
    
    \node[align=left, anchor=west] (wpo) at (-3, 6.5) {\footnotesize{SECI UART and GCI-GPIOs} \\ \textcolor{darkred}{\scriptsize{External output used in the evaluation board debugging setup.}}};
    \filldraw[fill=darkred!10,draw=darkred!50, align=center](-1,4.75) rectangle node (gci) {\footnotesize{GCI}} ++(1.5,0.4);
    \draw[<->] (-0.25,5.15)--(-0.25,6);
    \draw[-,gray,dashed] (-2.6,5.55) -- (2,5.55);
    
    \draw[->,color=darkred] (-2.9,4.1)--(-2.9,4.95)--(-1,4.95);
    \draw[->,color=darkred] (2.3,4.1)--(2.3,4.95)--(0.5,4.95);

    
    \path[->,color=darkred] (-2.9,3.4) edge node[sloped, anchor=left, above, text width=4.25cm,yshift=0.05em] {\footnotesize{WLAN RAM Sharing} \\ \vspace{0.5em} \tiny{\cveh: Information Disclosure}\\ \vspace{-0.75em} \tiny{\cveg: Code Execution}} (2.3,3.4);
    \path[->,color=darkred] (-2.9,3.0) edge node[sloped] {} (2.3,3.0);
    
    \path[<-] (-2.9,2.05) edge node[sloped, anchor=left, above, text width=4.25cm,yshift=-0.15em] {\footnotesize{WLAN/BT Access}} (2.3,2.05);
    \path[->] (-2.9,1.85) edge node[sloped] {} (2.3,1.85);
    
    \path[<->,color=darkred] (-2.9,0.1) edge node[sloped, anchor=left, above, text width=4.25cm,yshift=0.05em,xshift=1.65cm] {\footnotesize{GCI Coex I/F} \\ \vspace{0.5em} \tiny{\cvec: Denial of Service} \\ \vspace{0.5em}  \tiny{\cvef: Denial of Service}\\ \vspace{-0.75em} \tiny{\cvei: Information Disclosure}} (-1,0.1);
    \path[<->,color=darkred] (0.5,0.1) edge node[sloped] {} (2.3,0.1);
    
    \path[<->] (-2.9,-0.8) edge node[sloped, anchor=left, above, text width=4.25cm,yshift=-1.15em] {\footnotesize{Shared LNA Control} \\ \footnotesize{and Other Coex I/Fs}} (2.3,-0.8);
    
    \filldraw[fill=gray!10,draw=gray!50, align=center](-3.3,-3.7) rectangle node (clb) {\footnotesize{CLB}} ++(1.5,0.4);
    \draw[->] (-4.5,-3)--(-4.5,-3.5)--(-3.3,-3.5);

    \filldraw[fill=gray!10,draw=gray!50, align=center](2.2,-3.7) rectangle node (fem) {\footnotesize{FEM or SP3T}} ++(2,0.4);
    \node[align=left, anchor=west] (gh) at (3.2,-3.9) {\footnotesize{2.4\,GHz}};
    \filldraw[fill=gray!10,draw=gray!50, align=center](4.6,-3.7) rectangle node (spdt) {\footnotesize{FEM or SPDT}} ++(2,0.4);
    \node[align=left, anchor=west] (gh) at (5.6,-3.9) {\footnotesize{5\,GHz}};
    
    \draw[<->] (3.2,-3.7)--(3.2,-4.2)--(3.9,-4.2)--(3.9,-4.6);
    \draw[<->] (5.6,-3.7)--(5.6,-4.2)--(5.1,-4.2)--(5.1,-4.6);

    \draw[->] (-3.5,-2.05)--(-3.5,-2.5)--(3.1,-2.5); 

    \draw[-,gray,dashed] (3.4,-4.35) -- (5.4,-4.35); 
    \filldraw[fill=gray!10,draw=gray!50, align=center](3.4,-5) rectangle node (fem) {\footnotesize{Diplexer}} ++(2,0.4);
    \fill[fill=white,draw=black,yshift=-3.5cm,xshift=12cm] (-4,0)--(-3.6,0)--(-3.8,-0.2)--(-4,0);
    \draw[](5.4,-4.8)--(8.2,-4.8)--(8.2,-3.7);


	\end{tikzpicture}
\caption{Coexistence interfaces as documented for the \emph{Google Nexus 5} chip~\cite{bcm4339}, discovered vulnerabilities marked in \textcolor{darkred}{red}.}
\label{fig:coex_datasheet}
\end{figure*}
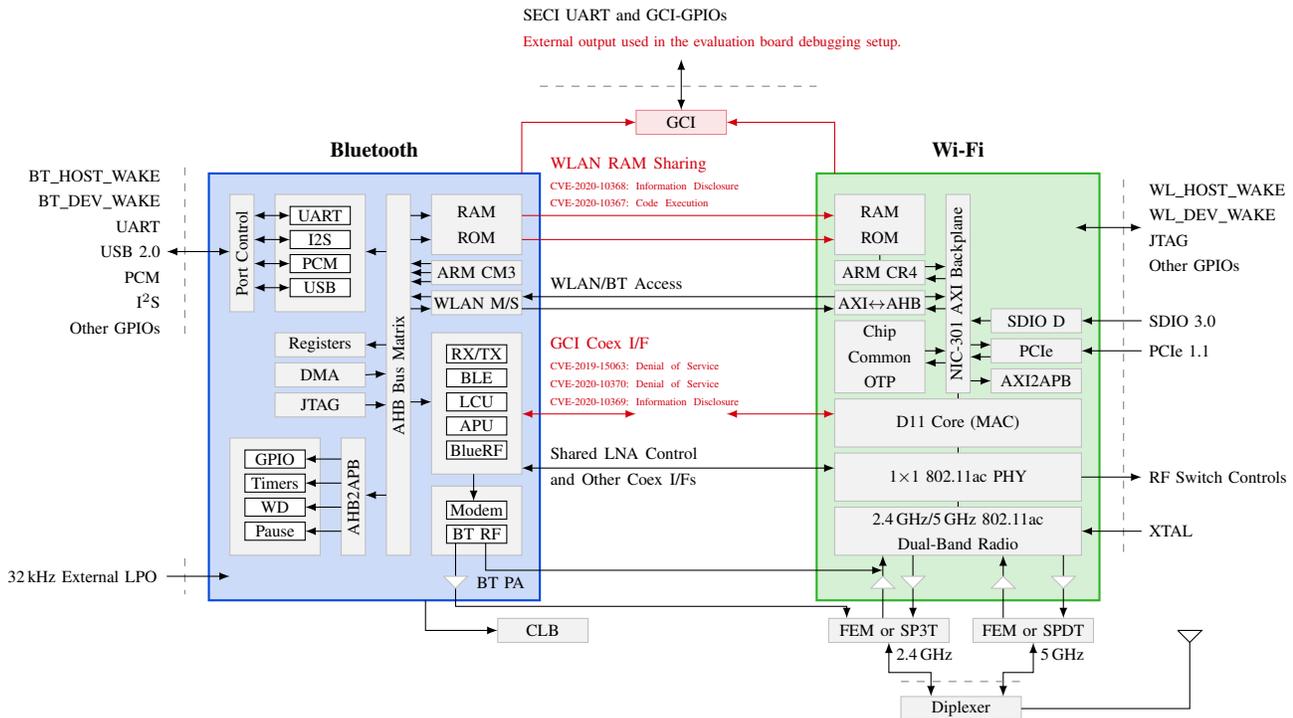

\section{Practical Coexistence Attacks}
\label{sec:practial}

We implement coexistence attacks on multiple chip types, components, and protocols.
This demonstrates that coexistence is exploitable in practice. 
First, we explain standardized coexistence mechanisms (\autoref{ssec:coexstandard}).
Then, we outline methods for finding coexistence mechanisms within proprietary chips for
analysis (\autoref{ssec:bcmexample}). Based on this knowledge, we exploit
a proprietary memory sharing mechanism (\autoref{ssec:archrev}), a proprietary coexistence protocol (\autoref{ssec:bcmcoex}),
as well as a standardized serial protocol (\autoref{ssec:silabs}).

%
%
%
%
%

\subsection{Standardized Coexistence Mechanisms}
\label{ssec:coexstandard}

As a first step for testing coexistence implementations,
we need to understand which chips implement which coexistence mechanisms
and if these could be vulnerable.
Two hardwired inter-chip coexistence mechanisms are both openly standardized and actually 
used by wireless chips or mentioned in their datasheets. These mechanisms are called \acf{PTA} and \acf{MWS}.


\subsubsection{Packet Traffic Arbitration}
\label{sssec:bgpta}
\ac{PTA}, specified by IEEE 802.15.2, addresses the coexistence of a co-located Wi-Fi and Bluetooth transmitter~\cite{pta}.
Based on timing and packet type information of both Wi-Fi and Bluetooth, it prevents potential
collisions and assigns priorities by packet types.
For example, Bluetooth 
voice
transmissions and Wi-Fi reception acknowledgment frames are given priority.
The standard defines three collaborative coexistence mechanisms
between \mbox{Wi-Fi} and Bluetooth~\cite{pta}: \emph{i)} A frequency notch filter on the physical layer, 
\emph{ii)} traffic type prioritization called \ac{PTA}, and \emph{iii)} time-based transmission division called \ac{AWMA}
on the MAC layer. All of these mechanisms can be combined. The standard does not propose
any \ac{DoS} mitigation.

\ac{PTA} was specified in 2003, but is still mentioned in recent wireless combo chip datasheets~\cite{coexcyw,coexmarvell,silabspta}.
\emph{Silicon Labs} chips support a basic \ac{PTA} mode without proprietary
extensions, which we will analyze later in \autoref{ssec:silabs}.


\subsubsection{Mobile Wireless Standards}
The Bluetooth specification defines \ac{MWS}, which targets Wi-Fi and LTE~\cite[p. 290ff]{bt52}.
\ac{MWS} defines inter-chip logical signals, which indicate packet priorities~\cite[p. 3227ff]{bt52}.
The logical coexistence signals are exchanged using \ac{UART}.
Bluetooth indicates priority receptions 
and, thus, can request that the {LTE}
chip ceases its transmission or refrains from starting a transmission.
Even though \ac{MWS} is part of the Bluetooth specification, it also considers
chips that integrate \mbox{Wi-Fi} functionality. Since such chips know
the \mbox{Wi-Fi} state within the \SI{2.4}{\giga\hertz} frequency band, there is also
a 
\mbox{Wi-Fi} priority signal. According to the specification, it is not
guaranteed that the {LTE} chip honors these priority signals. However,
if it does, this enables \ac{DoS} attacks.
If vendor-specific messages are used~\cite[p. 3252]{bt52}, information
disclosure side channels might also be possible.

In general, \emph{Broadcom} and \emph{Cypress} Bluetooth chips as well as \emph{Intel} LTE chips implement \ac{MWS}  command handlers.
We confirm that
\ac{MWS} is used on recent \emph{iPhones} with \emph{Intel} baseband chips.
\ac{MWS} messages can be found in \emph{PacketLogger} traces
as well as the system log~\cite{ioslogging}, for example, on the \emph{iPhone 7}, \emph{8}, \emph{11}, and \emph{SE2}.

\subsection{Identifying and Testing Coexistence Features in Practice}
\label{ssec:bcmexample}

In the following, we use a real-world example to illustrate how coexistence interfaces
are implemented and how to analyze them for security issues.
Most datasheets contain little to no information about coexistence features.
Thus, we focus on the previously confidential datasheet of the slightly outdated \emph{BCM4339}
Bluetooth/Wi-Fi combo chip by \emph{Broadcom}. The datasheet was released upon acquisition by
\emph{Cypress}~\cite{cypressbroadcom}.
We highlight the most important parts of this
datasheet in \autoref{fig:coex_datasheet}. In general, any shared component or connection
between the two cores of the combo chip could be attacked:
\begin{itemize}
\item \ac{GCI}, which is mentioned to be similar to \ac{PTA} but with vendor-specific extensions,
\item unidirectional WLAN RAM sharing,
\item bidirectional WLAN/Bluetooth access, with the exact purpose being undocumented,
\item \ac{LNA} control,
\item \ac{FEM} or \ac{SP3T} switch circuit connecting the antenna.
\end{itemize}
The datasheet also includes ``other'' coexistence interfaces without any further specification. All interfaces are undocumented, e.g., commands
and addresses to interact with them are missing.
The most interesting features are \ac{GCI}, which is protocol-based and can also be addressed externally,
and WLAN RAM sharing, which could potentially be abused for escalating between cores.
Further features, such as \ac{LNA} control and antenna connections, are likely only susceptible to \ac{DoS}
attacks. Thus, we only focus on the first two features. 

Proprietary coexistence interfaces in the \emph{Broadcom} chip are not meant to be
reprogrammed by external developers. Nonetheless, this is possible by using \emph{Nexmon}
for the Wi-Fi core~\cite{nexmon:project} and \emph{InternalBlue} for the Bluetooth core~\cite{mantz2019internalblue}.
These two projects provide a patching framework and contain many resources assisting further
reverse engineering.


\begin{table*}[tp]
\renewcommand{\arraystretch}{1.3}
\caption{Wi-Fi code execution and data leak through Wi-Fi/Bluetooth shared RAM (\cveh, \cveg).}
\label{tab:wlcrash}
\centering
\scriptsize
\begin{tabular}{|l|l|l|r|c|l|}
\hline
\textbf{Chip} & \textbf{Device} & \textbf{Tested OS with FW Updates} & \textbf{FW Build Date} & \textbf{FW Accesses \path{wl_buff}} & \textbf{Code Execution}\\
\hline
BCM4335C0 & Nexus 5 & Android 6.0.1 & Dec 11 2012 & $\times$ & \textcolor{gray}{?} \\
BCM4345B0 & iPhone 6 & iOS 12.4--12.5.1 & Jul 15 2013 & $\times$ & \textcolor{darkred}{$\pmb{\checkmark \bigstar}$}\\
BCM43430A1 & Raspberry Pi 3 & Raspbian Buster & Jun 2 2014 & $\times$ & \textcolor{gray}{?}\\
BCM4345C0 & Raspberry Pi 3B+/4 & Raspbian Buster & Aug 19 2014 & \textcolor{darkred}{$\pmb{\checkmark}$} & \textcolor{darkred}{$\pmb{\checkmark}$}\\
BCM4358A3 & Samsung Galaxy S6, Google Nexus 6P & Lineage OS 14.1 & Oct 23 2014  & $\times$ & \textcolor{darkred}{$\pmb{\checkmark \bigstar}$} \\
BCM20703A2 & MacBook Pro 2016 & -- & Oct 22 2015 & \textcolor{darkred}{$\pmb{\checkmark}$} & \textcolor{darkred}{$\bullet$} \\
BCM4355C0 & iPhone 7 & iOS 13.3--14.3 & Sep 14 2015 & $\times$ & \textcolor{darkred}{$\pmb{\checkmark}$} \\
BCM4347B0 & Samsung Galaxy S8/S8+/Note 8 & Android 8.0.0 & Jun 3 2016 & \textcolor{darkred}{$\pmb{\checkmark}$} & \textcolor{darkred}{$\pmb{\checkmark \bigstar}$} \\
BCM4347B1 & iPhone 8/X/XR & iOS 13.3--14.7 & Oct 11 2016 & \textcolor{darkred}{$\pmb{\checkmark}$} & \textcolor{darkred}{$\pmb{\checkmark}$} (\textcolor{darkred}{$\pmb{\bigstar}$} $<$14.3)\\
BCM4375B1 & Samsung Galaxy S10/S10e/S10+ & Android 9 & Apr 13 2018 & $\times$ & \textcolor{darkred}{$\pmb{\checkmark \bigstar}$} \\
BCM4375B1 & Samsung Galaxy S10/S10e/S10+/S20/Note 20 5G & Android 10 & Apr 13 2018 & $\times$ & \textcolor{darkred}{$\pmb{\checkmark}$} \\
BCM4375B1 & Samsung Galaxy S20/Note 20 5G & Android 11 & Apr 13 2018 & $\times$ & \textcolor{darkred}{$\pmb{\checkmark}$} \\
BCM4377B3 & MacBook Pro+Air, 2019--2020 (PCIe) & macOS 10.15.1--10.15.7 & Feb 28 2018 & $\times$ & \textcolor{darkred}{$\pmb{\checkmark}$}  (\textcolor{darkred}{$\pmb{\bigstar}$} $<$10.5.7)\\
BCM4364B3 & MacBook Pro+Air, 2019--2020 (UART) & macOS 10.15.4--10.15.6 & May 9 2018 & $\times$ & \textcolor{darkred}{$\pmb{\checkmark}$} \\
BCM4378B1 & iPhone 11/SE2 & iOS 13.3--13.5 & Oct 25 2018 & $\times$ & \textcolor{darkred}{$\pmb{\checkmark}$} \\
\hline
\end{tabular}
\caption*{\normalfont\scriptsize{
\hspace*{3.1em}\hspace*{1.07cm}\textcolor{gray}{?}\hspace*{0.08cm} Mentioned in datasheet but probably different mapping, did not work in our test. \\
\hspace*{3.1em}\hspace*{1.13cm}\textcolor{darkred}{$\bullet$}\hspace*{0.07cm} Likely vulnerable but no physical device available for testing. \\
\hspace*{3.1em}\hspace*{1.13cm}\textcolor{darkred}{$\pmb{\checkmark}$}\hspace*{0.0cm} Code execution within Wi-Fi successfully tested. \\
\hspace*{3.1em}\hspace*{1.1cm}\textcolor{darkred}{$\pmb{\bigstar}$}\hspace*{0.05cm} Kernel panic observed on the \ac{OS} (Android/iOS/macOS).\\ \\
\vspace*{-0.5em}\hspace*{3.1em}\normalfont{\hspace*{1.32cm} Issues persist on all tested up-to-date devices, but not all \ac{OS} updates were tested due to limited device and jailbreak availability.}}}
\end{table*}

\subsection{WLAN RAM Sharing (Broadcom \& Cypress)}
\label{ssec:archrev}

In the following, analyze the WLAN RAM sharing feature and find that
the Bluetooth core can abuse this feature for reading security-sensitive data
from the Wi-Fi core and reliably gain code execution.

\subsubsection{Reverse Engineering}

We recover the \emph{``WLAN RAM Sharing''} internals
using leaked symbols from \emph{Cypress WICED Studio 6.2}, a public wireless chip development platform by \emph{Cypress}~\cite{wiced, polypyus}.
It contains symbol information about the 
Bluetooth part of the \emph{BCM20703A2} combo chip, embedded into 2015--2016 \emph{MacBooks}.
In these symbols, we find a few functions following the naming scheme \path{wlan_buff_*}.
Based on these function names and a ROM dump from the \emph{MacBook}, we reverse-engineer the Bluetooth$\rightarrow$\mbox{Wi-Fi}
memory mapping.
When \mbox{Wi-Fi} is turned off, this memory area is all zeros. Otherwise---irrespective of whether Wi-Fi is currently connected to a network---this memory region is mapped.
As detailed in Appendix \ref{ssec:appendixmapping}, Bluetooth can read and write data within a large memory region.
The shared RAM address mapping is the same on all chips introduced since 2014.

The shared RAM region seems to be a \textbf{legacy feature}.
Bluetooth firmware compiled after 2016 does not cross-reference the shared RAM,
as listed in \autoref{tab:wlcrash}. The few symbol leaks let us assume that the shared RAM
was introduced for audio buffer sharing. It remains unclear why such a feature would require access
to almost the full Wi-Fi RAM region via Bluetooth and why it was never removed in
newer chips.

\subsubsection{RAM Sharing Code Execution}

Bluetooth can write to the Wi-Fi shared RAM area.
We confirm that this is sufficient for Bluetooth to execute code on the Wi-Fi core.
Due to its firmware patching mechanism, Wi-Fi executes 
code in writable memory regions~\cite{nexmon:project}. There are two suitable approaches to identify \mbox{Wi-Fi} RAM
areas that lead to code execution:
\begin{itemize}
\item Statically reverse-engineering Wi-Fi patches provided by the vendor, which are applied to the RAM region,
and find a location that is executed regularly, or
\item dynamically writing pre-defined chunks to the Wi-Fi RAM region until they appear in crash logs.
\end{itemize}

The Wi-Fi firmware is always loaded to the same addresses. Thus, once code execution
was confirmed on a single chip and firmware patch level, the identified region is valid on
all devices of the same model and patch level. 

We confirm code execution by analyzing crash logs. However, after initial confirmation and
exploit development, logging is no longer required.
\emph{Android}, \emph{macOS} and \emph{iOS} generate Wi-Fi crash logs, including a full chip RAM dump.
We use RAM dumps to confirm that we can write to the shared RAM and got code execution.
On \emph{Android}, \mbox{Wi-Fi} crash logs and memory dumps are written to \path{/data/vendor/log/wifi/}.
On some models, memory dumps are disabled, but the \mbox{Wi-Fi} chip still logs 
to the kernel, observable via \path{/dev/kmsg}.
\emph{macOS} and \emph{iOS} use a different debug format.
On \emph{iOS}, creating these logs requires an additional Wi-Fi debug profile~\cite{ioslogging}.
\emph{macOS} and \emph{iOS} both write to the common log directory. The folder
name already contains the crash cause and the file \path{SoC_RAM.bin} contains
the RAM dump.

Finding exploitable regions can be automated by sending randomized branch instructions
as bytecode. For example, if the Bluetooth chip writes the instruction \texttt{b 0xcafebabe} 
to a shared RAM address and Wi-Fi executes it, it will crash since there is no valid
code at the branch target. The Wi-Fi firmware fault handler 
forwards the exception to the Wi-Fi driver prior to resetting. Thus, the address
\texttt{0xcafebabe} appears in the \ac{OS} \mbox{Wi-Fi} crash logs.
The last address byte differs due to ARM-specific branch handling.
We find shared RAM regions
that \textbf{immediately
trigger \mbox{Wi-Fi} chip code execution}.

While using this method to identify executable code regions, the Bluetoth chip frequently continues sending bytes to the shared RAM region even after the Wi-Fi chip crashes. The Bluetooth chip can already control the shared RAM
during early Wi-Fi chip initialization. In this state, the Wi-Fi chip is not
connected to any network.
After each chip crash, the code execution finder should wait at least \SI{10}{\second},
ensuring that executable regions are reached during a regular chip state.

The Wi-Fi crash logs contain further insights into the chip's current
state and information. For example, when the chip crashes while being connected to
a network, the RAM dump contains the network name and password.
Thus, an attacker is able to access this information.

This attack only affects combo chips. Separate chips lack the bus used for memory sharing, and thus,
coexistence coordination via pure serial interfaces 
is not affected.
Almost all chips manufactured by \emph{Broadcom} are combo chips.
Some \emph{Cypress}-branded chips, such as in the \emph{Raspberry Pi}, are similarly affected by shared RAM attacks~\cite{cypressspectra}.

\subsubsection{Over-the-Air Code Execution}

Bluetooth$\rightarrow$Wi-Fi code execution can be triggered over the air. 
We show the potential of coexistence exploitation by building an over-the-air \ac{PoC} on a \emph{Samsung Galaxy S8}.
The authors of \emph{Frankenstein} published CVE-2019-11516~\cite{frankenstein}, an over-the-air Bluetooth code execution vulnerability.
This issue was reported in April 2019, and patches were rolled out in fall 2019. We downgrade the \emph{S8} Bluetooth firmware to a
January 2019 patchlevel to ensure it is still vulnerable. We keep the remaining system on an up-to-date \emph{Samsung} stock ROM,
including the Wi-Fi firmware.
The following equipment is required to perform the attack:

\begin{itemize}
\item \emph{CYW20735} evaluation board with a firmware modification to set shellcode device names and change the Bluetooth MAC address,
taken from the \emph{Frankenstein} repository~\cite{frankenstein-eir}.
\item \emph{Raspberry Pi} or other \emph{Linux} device with the \emph{BlueZ} stack.
\item A battery pack (optional).
\end{itemize}

\textbf{The total costs of this portable setup are below USD 100.}
Such a mobile attacker is hard to locate.
While the original \ac{PoC}
was meant to write arbitrary memory on the Bluetooth chip, combining this with the shared Wi-Fi RAM
allows us to control arbitrary Wi-Fi memory, resulting in \mbox{Wi-Fi} code execution.
On the specific Wi-Fi version running on the \emph{Samsung Galaxy S8}, the Bluetooth address
\texttt{0x6841d2} is mapped to the Wi-Fi address \texttt{0x1841d2}, and the code at this address is executed regularly.
When writing the shellcode \texttt{b 0xcafebabe} to this address via Bluetooth, the Wi-Fi console output read from \path{/dev/kmsg} shows
that the code is executed:

\begin{lstlisting}
CONSOLE: 000000.454 wl1: Broadcom BCM4361 802.11 Wireless Controller 13.38.63 (B0 Network/rsdb)
CONSOLE: 000000.456 ThreadX v5.6 initialized
CONSOLE: TRAP 3(2bfea0): @\textcolor{darkred}{pc cafebabc}@, lr 1843ef, sp 2bfef8, ...
\end{lstlisting}

Note that \emph{CVE-2019-11516} was only fixed on devices that still received official security updates in fall 2019.
Devices like the \emph{Google Nexus 5} or \emph{Samsung Galaxy S6} remain unpatched. Unofficial \emph{Android} images, e.g., \emph{LineageOS},
only contain \ac{OS} updates---updated firmware patches require a collaboration between \emph{Broadcom} and, respectively, \emph{Samsung} and \emph{Google}.

\subsubsection{Wi-Fi Kernel Driver Issues}

Finding Wi-Fi kernel vulnerabilities is not a focus of this paper and has conceptually been covered by previous work~\cite{2020:googleprojectzero,2017:googleprojectzero,yuwangusa}.
Yet, our Bluetooth$\rightarrow$Wi-Fi coexistence test setup triggers kernel panics across multiple \acp{OS}.
After our reports, the vendors fixed kernel panics caused by missing PCIe and \mbox{Wi-Fi} state management.
Kernel panics and fixes are listed in \autoref{tab:wlcrash}.

All these crashes were produced while probing the Bluetooth$\rightarrow$Wi-Fi shared RAM interface. This indirectly fuzzes the Wi-Fi$\rightarrow$host interface, most of the time
by crashing the Wi-Fi chip when the \ac{OS} driver does not expect it.
Thus, the majority of these kernel panics are caused by PCIe bus timeouts and failed attempts to bring up the Wi-Fi chip again on \emph{Android}, \emph{macOS}, and \emph{iOS}.
Interestingly, some of these crashes indicate more substantial issues---e.g., a malformed PCIe \ac{IOMMU} request
on \emph{iOS}. This means that the fuzzer also manipulates data sent to the host via PCIe.

Overall, the stability issues found in the kernels indicate that \mbox{Wi-Fi} drivers were not tested sufficiently.
We assume that especially unexpected states that over-the-air setups cannot trigger but via coexistence
were not considered, such as invalid replies during driver initialization or network connection setup.
While we were working on coexistence issues inside the chips, other researchers found various issues within \emph{Apple}'s Wi-Fi
stack in parallel~\cite{2020:googleprojectzero,yuwangusa}.


\subsection{Serial Enhanced Coexistence (Broadcom \& Cypress)}
\label{ssec:bcmcoex}

The proprietary \acf{SECI} by \emph{Broadcom} and \emph{Cypress} is used internally
in combo chips. In addition, it is exposed on chips that only support Bluetooth
or Wi-Fi, and it can be manually connected.
\ac{SECI} is largely undocumented,
except for the information covered in \autoref{sssec:doc}.
Its functionality can be observed with a logic analyzer as described in \autoref{sssec:reversing},
resulting in understanding the \ac{SECI} physical layer in \autoref{ssec:phy}. 
After understanding the protocol, we can mount \ac{DoS} and information
disclosure attacks from the firmware---without a logic analyzer or other hardware modifications.
A Bluetooth \ac{DoS} on the grant reject scheme is demonstrated in \autoref{ssec:wifidos}.
A more severe attack, where Wi-Fi observes packet types and timings of Bluetooth keyboards
to determine keystroke timings, is shown in \autoref{sssec:hid}.

\subsubsection{Documentation}
\label{sssec:doc}

The \emph{BCM4339} datasheet in \autoref{fig:coex_datasheet} shows multiple coexistence interfaces~\cite{bcm4339}.
\ac{GCI} supports various coexistence mechanisms, such as standardized \ac{MWS}
and proprietary \ac{SECI}.
\ac{SECI} uses \ac{UART} to transmit \SI{64}{\bit} coexistence data~\cite{coexcyw, coexseci}.
\mbox{Wi-Fi} sends its current transmit channels to Bluetooth, which blocklists them.
Moreover, \ac{SECI} contains timeout parameters, such as \ac{ACL} and \ac{SCO} timeout limits, powersave and idle timers, as well as medium request and grant timers.
Priorities of Wi-Fi and Bluetooth are hardcoded. Audio, video, \ac{BLE}, and \acp{HID} have the highest priority, while file transfer has the lowest priority.
During Wi-Fi powersave, all Bluetooth requests are granted.
Otherwise, coexistence polls between \mbox{Wi-Fi} and Bluetooth.

These features extend \ac{PTA}. As the \emph{Cypress} coexistence application note states,
the chips already implement \ac{PTA}~\cite{coexcyw}. \ac{SECI} augments \ac{PTA} with additional information to enable more
advanced coexistence methods.

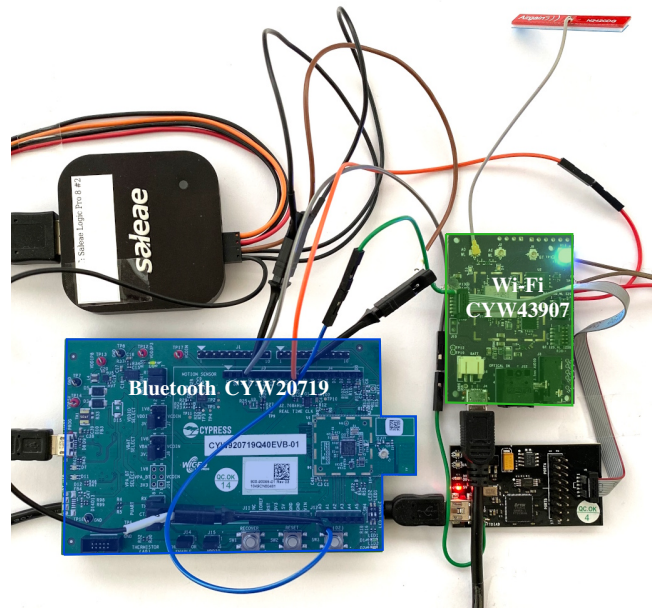
\begin{figure}[!b]
	\vspace{-2em} 

	\begin{tikzpicture}[minimum height=0.55cm, scale=0.8, every node/.style={scale=0.8}, node distance=0.7cm]
 
	\node[inner sep=0pt] (bt) at (0,0)
{\includegraphics[width=1.2\linewidth,bb=0 0 1024 994]{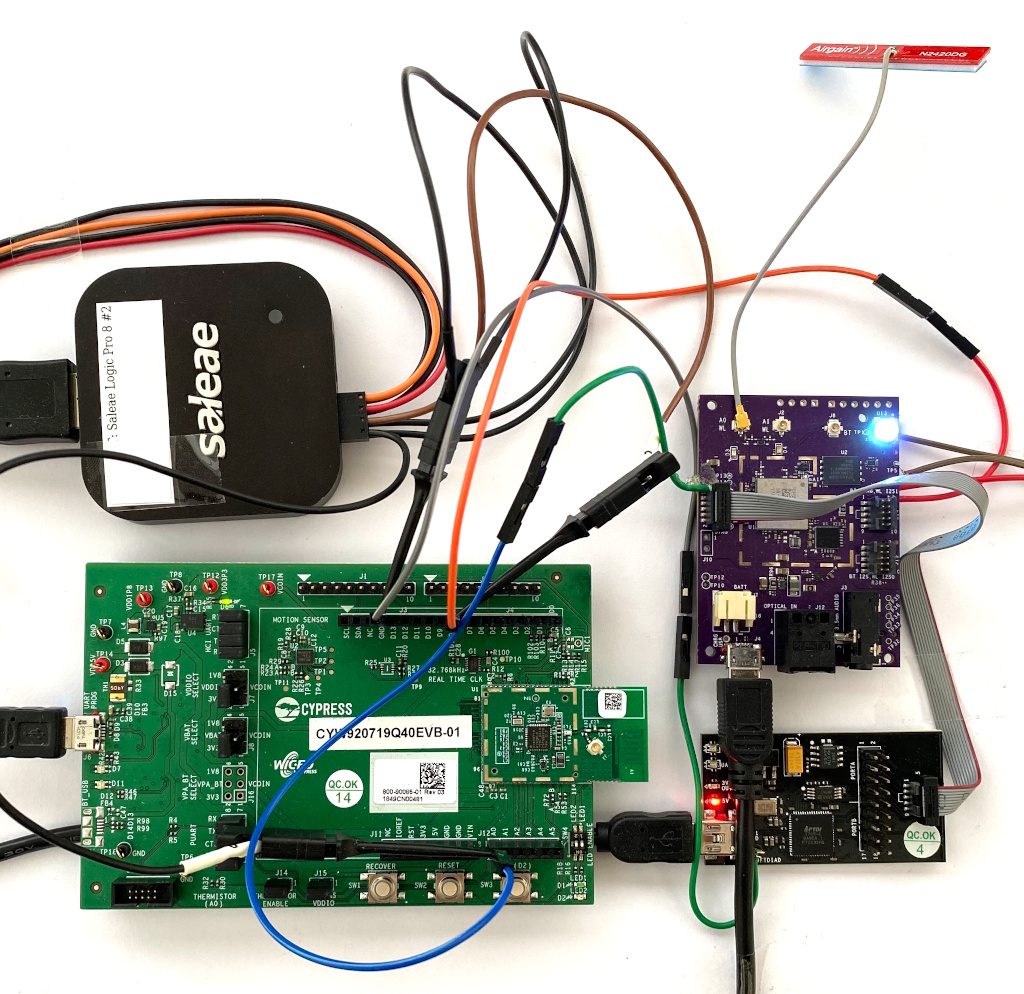}};	
    \fill[fill=darkblue,draw=darkblue,thick,fill opacity=0.35] (-4.45,-4.28) -- (-4.4,-0.7) -- (0.8,-0.7) -- (0.82,-1.98) -- (1.4,-1.98) -- (1.4, -2.95) -- (0.84,-2.95) -- (0.85,-4.2) -- (-4.45,-4.28);
    \node[anchor=west,white] (chiptxt) at (-3.5,-1.5) {\textbf{Bluetooth CYW20719}};
    \fill[fill=darkgreen,draw=darkgreen,thick,fill opacity=0.35] (1.9,1.05) -- (4,1.05) -- (4,-1.8) -- (1.9,-1.75) -- (1.9,1.05);
    \node[anchor=west,white,align=center] (chiptxt) at (2.1,0) {\textbf{Wi-Fi} \\ \textbf{CYW43907}};
	\end{tikzpicture}
\caption{Debugging setup for the \emph{Cypress} \ac{SECI} between a separate Wi-Fi and Bluetooth chip.}
\label{fig:cypress}
\end{figure}

\subsubsection{Reverse-engineering}
\label{sssec:reversing}

\ac{SECI} uses a Bluetooth$\rightarrow$Wi-Fi wire and a \mbox{Wi-Fi}$\rightarrow$Bluetooth wire.
Both wires operate at \SI{3}{MBaud} and indicate all packets, including metadata, such as the packet type. For comparison, the Bluetooth$\leftrightarrow$Host UART interface that carries all Bluetooth data also transmits with \SI{3}{MBaud}.
Both the Wi-Fi firmware and \emph{D11} core as well as the Bluetooth firmware access \ac{SECI}.
The Wi-Fi \emph{D11} core is a reprogrammable low-level processor that implements a state machine for the
first packet processing stage~\cite{nexmon:project}.

Combo chips have an integrated \ac{SECI} connection, but even
separate \mbox{Wi-Fi} and Bluetooth modules can be connected
with \ac{SECI}.
Each chip generation implements a slightly different 
version, and only specific chips can be connected~\cite{coexcyw}.
Major differences are \SI{48}{\bit} or \SI{64}{\bit} data size and \SI{3}{MBaud} or \SI{4}{MBaud} data rate.
Irrespective of physical layer details,
\ac{SECI} is abstracted as hardware-mapped registers.
Once a fresh value is written to an output register on one
wireless core, it is received as input on the other wireless core.

\begin{figure*}[tp]

\begin{subfigure}[c]{1.0\textwidth}

\includegraphics[width=1.0\textwidth]{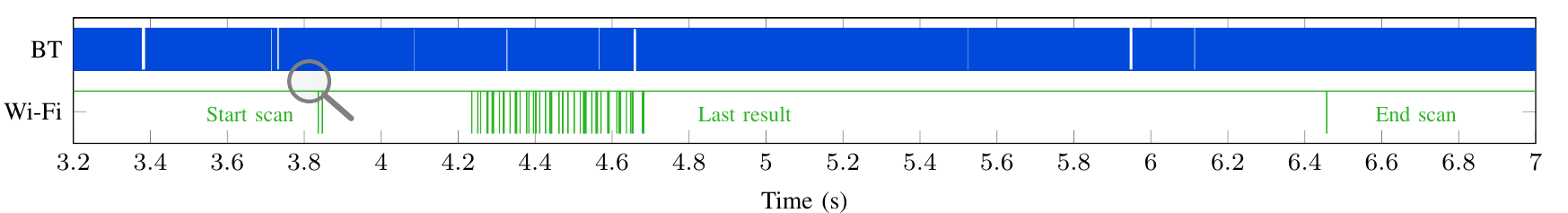}

\vspace{-0.3em} 
\subcaption{SECI while streaming music via Bluetooth and scanning for Wi-Fi access points.}
\label{sfig:seci_phy1}
\end{subfigure}

\begin{subfigure}[c]{1.0\textwidth}
\includegraphics[width=1.0\textwidth]{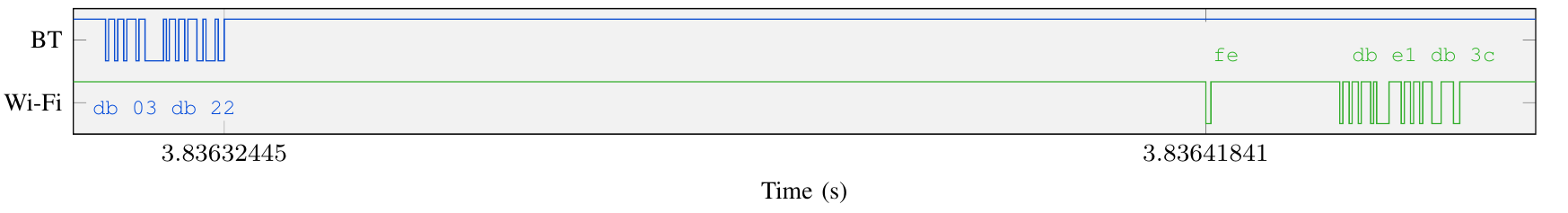}

\vspace{-0.5em} 
\subcaption{SECI while streaming music via Bluetooth and scanning for Wi-Fi access points, zoomed in to first Wi-Fi peak.}
\label{sfig:seci_phy2}
\end{subfigure}

\vspace{-1em} 
\caption{Initial SECI observations.}
\label{fig:seci_phy}
\end{figure*}

For the initial analysis, we reproduce a coexistence setup with a separate \mbox{Wi-Fi} and Bluetooth
chip~\cite{coextp}.
Then, we intercept the exposed \ac{SECI} with a logic analyzer, as shown in \autoref{fig:cypress}.
Wi-Fi \texttt{TP17} and \texttt{TP18} are connected to Bluetooth \texttt{D8} and \texttt{A0}, respectively.
The black board is required for Wi-Fi programming and debugging but has no further wireless functionality. 
In this setup, all information exchanged between Wi-Fi and Bluetooth only  originates from \ac{SECI}. Antenna locks and other signal-related effects can be excluded.

With this setup, \ac{SECI} messages can be observed, and their meanings can be reverse-engineered.
\ac{SECI} is not implemented inside the firmware but an external hardware component.
The Bluetooth and \mbox{Wi-Fi} firmware only use hardware register mappings to write values
over the serial protocol (mapping details for reproducing our experiments are provided in Appendix \ref{ssec:appendixseci}).
According to our threat model, the attacker only has code execution on one of the
wireless core, but cannot attach cables. Thus, the attacker can only observe values in these registers at a maximum granularity of
fixed clock cycles.
We measure that the \ac{SECI} message timing has \SI{200}{\nano\second} standard deviation on the Wi-Fi \emph{D11} core.
This jitter is way below the \emph{D11} core \ac{SECI} sampling rate, meaning that
\textbf{the attacker cannot observe side-channels based on jitter.}

\subsubsection{Physical Layer Working Principle}
\label{ssec:phy}

Even though the high-level documentation states the basic features of \ac{SECI}
and that it is probably transmitting \SI{48}{\bit} or \SI{64}{\bit} blocks, we do
not know its exact working. Thus, we run an initial test series.
The Bluetooth input is supplied by a \emph{Linux} host that is decoupled from the \mbox{Wi-Fi} chip, 
such that \mbox{Wi-Fi} and Bluetooth can be measured independently.
On the \emph{Linux} host, we attach the \emph{CYW20719} evaluation board to \emph{BlueZ}
and then use it for audio streaming and keyboard input. Both use \ac{ACL} packets
for data transmission but at different rates.
The \mbox{Wi-Fi} evaluation board \ac{RTOS} provides a command-line interface
to join an access point as a station, acting as an access point itself or sending pings across the
network. \autoref{fig:seci_phy} shows two captures with this setup.

In the first capture (\autoref{sfig:seci_phy1}), Bluetooth streams music
while Wi-Fi scans for access points. The pattern indicating a scan has two peaks in the
beginning and one in the end. The intermediate peaks differ slightly depending on
the number of scan results. 
Each peak carries serial data (see \autoref{sfig:seci_phy2}).
The logic analyzer decodes a serial protocol.
Most protocol values start with \texttt{db}.
Each value indicates a certain action.
For example, when Wi-Fi starts scanning, the first peak decodes as
\textcolor{darkgreen}{\texttt{fe db e1 db 3c}}.
Each packet has a variable length, with a maximum of up to \SI{64}{\bit}.

\input{pics/grantrejectdos.tex}

\input{pics/hid.tex}

\subsubsection{Bluetooth Grant and Reject DoS (\cvef)}
\label{ssec:wifidos}

Once \ac{SECI} is set up, Bluetooth waits for grants to send packets.
In case \SI{2.4}{\giga\hertz} Wi-Fi is enabled, Wi-Fi might reject Bluetooth packets, as they operate on the same frequency band.
Note that Wi-Fi does not reject Bluetooth packets if it is connected to a \SI{5}{\giga\hertz} access point
or if it is disabled.

Wi-Fi can abuse this grant and reject pattern.
If Wi-Fi does not signal that it is inactive but also stops
sending grants, Bluetooth pauses sending packets containing data because its requests are rejected.
We build this as \emph{D11} core \ac{PoC} for the \emph{Nexus 5}
as well as the \emph{CYW439037+CYW20719} evaluation board.
For the latter, the \ac{PoC} can temporarily be enabled and disabled
by sending a Wi-Fi packet with a particular payload. As shown in \autoref{fig:grantrejectdos},
Wi-Fi pauses sending \ac{SECI} messages during this attack period.
In the depicted example, Bluetooth is streaming audio, but pauses in case where Wi-Fi does not send any further \ac{SECI}
messages.
If the period
only lasts a few seconds, Bluetooth is able to maintain active connections with keep-alive null packets
but stops sending data. Otherwise, connections time out.

\subsubsection{Inferring Keypress Timings (\cvei)}
\label{sssec:hid}

In what follows, we show that an attacker with control over the
Wi-Fi core can infer exact Bluetooth packet timings and their content type.
This can be used to determine if a packet contained data and, thus,
observe if a packet sent by a Bluetooth keyboard contains a keypress.
Even though we only demonstrate this for a keyboard, further information disclosure attacks based on 
metadata might be possible.

\paragraph{Keypress Attacker Model}

Keypress timings, as observable by an attacker, can be analyzed statistically.
This becomes interesting for inferring passwords and password lengths. For example, after a long idle time,
the user likely first enters login credentials.
Previous work has shown that timing attacks on keyboard-based input is possible and
developed working statistical methods~\cite{brokenstrokes, song2001timing}.
Thus, keypress timings should not leak via coexistence protocols.

\paragraph{Setup}

We pair a \emph{Linux} host with a \emph{CYW20719} Bluetooth chip with a keyboard.
On the \emph{Linux} host, we can record a ground truth about what is happening on the
Bluetooth interface using \emph{Wireshark}. Each keypress is represented as \acf{ACL} data.
Then, we compare these recordings to \ac{SECI}.

\paragraph{Observing Bluetooth via Wi-Fi}

We record traces of multiple \ac{HID} keyboards.
Typical \ac{HID} devices operate at \SI{15}{\milli\second}.
The minimum interval in classic Bluetooth is \SI{1.25}{\milli\second}~\cite[p.~2318]{bt52}.
In practice, we find keyboards
with timings of up to \SI{30}{\milli\second}, as shown in \autoref{tab:hidtiming}.
The Wi-Fi \emph{D11} core polls every \SI{1.25}{\milli\second} and is able to observe each  packet's metadata
at the same rate as Bluetooth packets are sent.

\begin{table}[!t]
\renewcommand{\arraystretch}{1.3}
\caption{Timings of Bluetooth \ac{HID} keyboards.}
\label{tab:hidtiming}
\centering
\scriptsize
\begin{tabular}{|l|r|}
\hline
\textbf{Product} & \textbf{Timing}\\
\hline
Apple Wireless Keyboard & \SI{12.5}{\milli\second} \\
Apple Magic Keyboard & \SI{15}{\milli\second} \\
Adafruit Mini Keyboard & \SI{30}{\milli\second} \\
\hline
\end{tabular}
\vspace{-1em} 
\end{table}

\ac{HID} devices like keyboards transfer their keystrokes using the \ac{ACL} protocol.
Thus, on the Bluetooth end, the \ac{ACL} coexistence handler \path{_ecsi_gci_HandlerACL}
is responsible. This handler writes values to the output register \path{gci_output}.
Once a value is written to this register, the Bluetooth chip sends it over the
\ac{SECI} interface. Then, this value appears in the Wi-Fi \emph{D11} core register.

We analyze which Bluetooth \ac{HID} events the Wi-Fi \emph{D11} core 
can capture. To this end, we hook the \path{_ecsi_gci_HandlerACL}. This enables us to observe
the \ac{SECI} values as written by the Bluetooth chip and align these with the
actual keystrokes as decoded by the Bluetooth chip. In parallel, we attach a logic
analyzer to \ac{SECI}, which is exposed on the \emph{CYW20719} evaluation board.
Even though we already know the values that will be sent to the Wi-Fi \emph{D11} core,
this enables us to observe potential jitter. The different outputs produced by this experiment
are depicted in \autoref{fig:keypress}.

Using this setup, we find that keystroke events are indicated by the value \texttt{85}
sent over the serial interface, while empty packets during inactivity are indicated
by the value \texttt{05}. Thus, the \textbf{Wi-Fi \emph{D11} core is able to distinguish
between packets with and without keystrokes}.
The granularity of frames sent over the \ac{SECI} physical layer is equal to the \ac{HID} device events, which is 
\SI{30}{\milli\second} in our example.
Moreover, we observe a jitter of \SI{200}{\nano\second}, which is way below the Wi-Fi \emph{D11} core poll interval, and, thus, cannot be captured by the Wi-Fi core. 


\subsection{Packet Traffic Arbitration (Silicon Labs)}
\label{ssec:silabs}

In the following, attacks on a \ac{PTA}-based interface are shown.
We analyze the \emph{Silicon Labs} \mbox{Wi-Fi}$\leftrightarrow$Bluetooth implementation.
\autoref{sssec:sipta} introduces the basic concept behind \ac{PTA}.
The ana\-lysis setup for the \emph{Silicon Labs} coexistence development kit is described in \autoref{sssec:sisetup}.
The plain \ac{PTA} implementation is susceptible to \ac{DoS} (see \autoref{sssec:sidosbt} and \ref{sssec:sidoswifi}), as expected according
to the \ac{PTA} specification. The \emph{Silicon Labs} implementation introduces a significant
jitter that is sufficient to let Wi-Fi infer basic Bluetooth protocol activities and vice versa (see \autoref{sssec:siinfbt} and \ref{sssec:siinfwifi}).

\subsubsection{PTA Working Principle}
\label{sssec:sipta}

As briefly outlined in \autoref{sssec:bgpta}, \ac{PTA} allows for prioritizing transmissions
like Bluetooth voice and \mbox{Wi-Fi} acknowledgments. The Bluetooth voice mode in headsets is also known as \ac{SCO},
which is optimized for low-latency voice, in contrast to music and data transmission. Since it is
low-latency, it should be prioritized. In Wi-Fi, each data frame is immediately acknowledged by the
receiver within a fixed time slot. If the acknowledgment is not received, the data frame must be retransmitted---causing significantly more congestion.

\ac{PTA} allows two wireless chips to coordinate their transmissions. Both ends get the
ultimate power to request a priority transmission. In an over-the-air scenario, if both ends 
send their data at once, it would collide, and both transmissions are lost. Thus, \ac{PTA}
decides which transmitter wins the competition and tells the other end to wait.
\ac{PTA} is a simple scheme that only allows to request
a transmission slot and optionally mark its priority. No metadata about packet types is
included. Each transmitter decides on its own what should be prioritized.

\subsubsection{Experimental Setup}
\label{sssec:sisetup}

As shown in \autoref{fig:sipta}, \ac{PTA} coexistence is based on three signals: \texttt{REQUEST} and \texttt{PRIORITY} sent by
Bluetooth and \texttt{GRANT} sent by Wi-Fi.
\emph{Silicon Labs} integrates the \ac{PTA} coexistence controller
into the Wi-Fi chip.
Thus, Wi-Fi knows whether it has a priority transmission
and schedules it without asking permission.
However, Bluetooth can request priority for urgent transmissions.

\emph{Silicon Labs} offers a separate coexistence development kit to be plugged into a Wi-Fi and a
Bluetooth development board. The coexistence kit provides \ac{PTA} and additional debug outputs. Parts of the documentation
are non-public, but a public application note explains the basic configuration
and logic analyzer outputs~\cite{silabspta}.

\tikzset{>=latex}
\begin{figure}[!h]

	\center
	\begin{tikzpicture}[minimum height=0.55cm, scale=0.8, every node/.style={scale=0.8}, node distance=0.7cm]

    \filldraw[fill=darkblue!20,draw=darkblue,thick](0,0) rectangle node (bt) {BRD4181B} ++(4,1.5);
    \node[anchor=north,align=center, above=of bt.north,yshift=-0.3cm] (chiptxtb) {\textbf{Bluetooth}};    
    
    \filldraw[fill=darkgreen!20,draw=darkgreen,thick, align=center](6,0) rectangle node (wf) {BRD4321A \\ \small{\emph{Integrated PTA Controller}}} ++(4,1.5);
    \node[anchor=north,align=center, above=of wf.north,yshift=-0.5cm] (chiptxtb) {\textbf{Wi-Fi}};

    \path[->,thick,darkblue] (4,0.75+0.5) edge node[sloped,yshift=0.2cm,align=center] {\texttt{REQUEST}} (6,0.75+0.5);
    \path[->,thick,darkblue!50!black] (4,0.75) edge node[sloped,yshift=0.2cm,align=center] {\texttt{PRIORITY}} (6,0.75);
    \path[<-,thick,darkgreen] (4,0.75-0.5) edge node[sloped,yshift=0.2cm,align=center] {\texttt{GRANT}} (6,0.75-0.5);


	\end{tikzpicture}
\caption{PTA implementation by \emph{Silicon Labs}.}
\label{fig:sipta}
\end{figure}
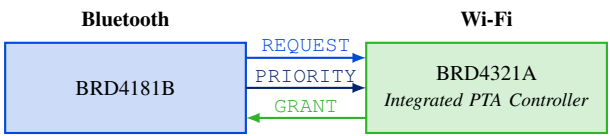

\subsubsection{Wi-Fi$\rightarrow$Bluetooth DoS}
\label{sssec:sidosbt}
In the first experiment, the Bluetooth development board is configured as \ac{BLE} beacon,
which regularly advertises its identity. 
The beacon
sends coexistence \texttt{REQUEST} signals during this time.
Once Wi-Fi changes the \texttt{GRANT} signal from \texttt{0} to \texttt{1} (inverted logic), the
requests sent by Bluetooth become shorter. Moreover, the beacon disappears in a
Bluetooth scanner app, meaning that it stops sending packets.
While this is the expected coexistence behavior, Bluetooth does not continue sending
\ac{BLE} advertisements, even if \texttt{GRANT} is set to \texttt{1} for multiple
minutes---causing a permanent \ac{DoS}.

Note that the \ac{PTA} configuration does not make any difference for the Bluetooth
side, since it only observes the resulting \texttt{GRANT} signal generated by
the \ac{PTA} controller.

\subsubsection{Bluetooth$\rightarrow$Wi-Fi DoS}
\label{sssec:sidoswifi}

Before turning on the Wi-Fi radio, five different scheduling priorities can be configured in the integrated \ac{PTA} controller
by setting them in the software development platform~\cite{silabsptaprios}.
The lowest priority for Wi-Fi and highest for the coexistence device, which is Bluetooth in this setup,
is \path{COEX_MAXIMIZED}. In this mode, Wi-Fi connections might be dropped. The opposite, 
\path{WLAN_MAXIMIZED}, maximizes Wi-Fi priority at the cost of Bluetooth.
The default configuration is \texttt{BALANCED}.

As summarized in \autoref{tab:siwifidos}, we test if any of these configurations 
are susceptible to \ac{DoS} attacks.
The \ac{PTA} priority changes how often
Bluetooth coexistence requests are granted. If Bluetooth makes excessive prioritized requests and is given
a priority by the \ac{PTA} configuration, this influences the \mbox{Wi-Fi} throughput. 
When configuring the Wi-Fi device as an access point and pinging it as a client, prioritized Bluetooth requests lead to \SI{100}{\percent} packet loss
for almost all settings. 
Even with the Wi-Fi priority set to \path{WLAN_HIGH}, packets are significantly delayed.
Only with \path{WLAN_MAXIMIZED} priority, 
a Bluetooth \texttt{PRIORITY} request cannot override the Wi-Fi priority.

\begin{table}[!h]
\renewcommand{\arraystretch}{1.3}
\caption{\emph{Silicon Labs} Wi-Fi DoS via Bluetooth requests.}
\label{tab:siwifidos}
\centering
\scriptsize
\begin{tabular}{|l|l|l|}
\hline
\textbf{PTA Priority} & \textbf{\texttt{REQUEST}, no \texttt{PRIO} set} & \textbf{\texttt{REQUEST}, \texttt{PRIO} set}\\
\hline
\path{COEX_MAXIMIZED} & DoS & DoS \\
\path{COEX_HIGH} & DoS & DoS \\
\path{BALANCED} & DoS & DoS \\
\path{WLAN_HIGH} & Grant periodically denied & DoS \\
\path{WLAN_MAXIMIZED} & Wi-Fi gets priority & Wi-Fi gets priority \\
\hline
\end{tabular}
\end{table}

\begin{figure}[!b]
	\begin{tikzpicture}[every node/.style={font=\footnotesize}]
		\begin{axis}
			[
			height=5cm, 
			width=\columnwidth, 
			boxplot/draw direction = y,
			xtick={1,3,5},
			xticklabels={Connected (Idle), Indication, Notification},
			ytick={-300,-200,-100,0,100,200,300},
			ylabel={{Time in \si{\micro\second}}}
			]
			\addplot+[darkred,
			boxplot prepared={
				draw position = 1,
				median=-12,
				upper quartile=3,
				lower quartile=-27,
				upper whisker=30,
				lower whisker=-190
			},
			] coordinates {};
			\addplot+[darkblue,
			boxplot prepared={
				draw position = 3,
				median=-171,
				upper quartile=-151,
				lower quartile=-183,
				upper whisker=226,
				lower whisker=-306
			},
			] coordinates {};
			\addplot+[gray,
			boxplot prepared={
				draw position = 5,
				median=-85,
				upper quartile=-69,
				lower quartile=-100,
				upper whisker=302,
				lower whisker=-286
			},
			] coordinates {};
		\end{axis}
	\end{tikzpicture}
	\caption{\texttt{REQUEST} offset compared to \SI{625}{\micro\second} Bluetooth slots.}
	\label{fig:bluetooth-timing}
\end{figure}
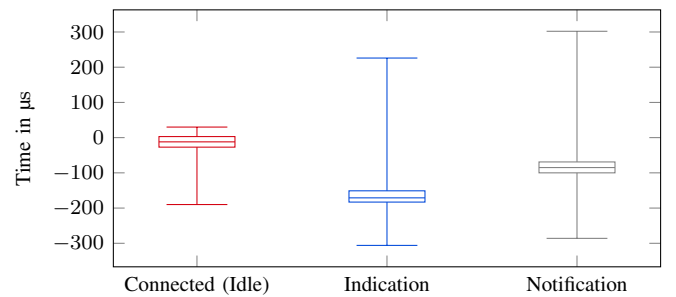

\subsubsection{Wi-Fi$\rightarrow$Bluetooth Traffic Type Disclosure}
\label{sssec:siinfbt}

The three \ac{PTA} wires are set to high or low, and do not contain information about \mbox{Wi-Fi} or Bluetooth packets. However, the \emph{Silicon Labs} implementation introduces a significant
jitter in Bluetooth \texttt{REQUEST} signals, which Wi-Fi can observe. According to the \emph{Silicon Labs}
documentation, coexistence signals are sampled at a rate of \SI{10}{\micro\second}. The measurement with
a logic analyzer is more precise. However, an attacker with Wi-Fi code execution is limited to this rate.
Interestingly, the jitter exceeds \SI{10}{\micro\second} by up to $+$\SI{302}{\micro\second}.
The jitter stays within $\pm$\SI{312.5}{\micro\second}, which is the maximum acceptable jitter for \ac{BLE} coexistence to work at all,
since \ac{BLE} uses \SI{625}{\micro\second} time slots.

\ac{BLE} traffic can be sent unacknowledged as notification
or acknowledged as indication~\cite[p. 1584ff]{bt52}.
As shown in \autoref{fig:bluetooth-timing}, there are major differences in slot offsets depending on the
traffic type. The interquantile range of the observed offsets stays within approximately \SI{30}{\micro\second}
for all types, but the median shifts drastically.

The jitter only leaves a very short reaction time for \mbox{Wi-Fi} to stay within one Bluetooth time slot from a coexistence performance perspective.
Jitter is not only a performance issue but also \textbf{\mbox{Wi-Fi} can
observe additional information about Bluetooth traffic types by approximating the jitter}.
This is surprising because the raw \ac{PTA} protocol does not contain any metadata.
However, the information extracted using the jitter is less fine-grained than
metadata from vendor-specific extensions.
Even though this measurement focuses on extracting information about packets sent by the Bluetooth chip,
further attacks like side-channels on the Bluetooth chip's calculation might be possible.
Yet, such attacks are likely prevented when restricting observations to the
\ac{PTA} sampling window of \SI{10}{\micro\second} on the Wi-Fi chip.

\subsubsection{Bluetooth$\rightarrow$Wi-Fi Activity Disclosure}
\label{sssec:siinfwifi}
In the opposite direction, the Wi-Fi \texttt{GRANT} signals also have significant time variances if 
Wi-Fi is under high load. However, we assume the same hardware-based sampling constraints for an attacker with
on-chip code execution. These prevent fine-grained measurements for effective side channels.

In our experiment, the Wi-Fi development board runs a minimal network stack, including a few testing tools.
It includes an \texttt{iperf} implementation that can measure TCP throughput.
An \texttt{iperf} client connects to the
Wi-Fi development board, configured as an access point, and sends TCP traffic with a
throughput of \SI{7}{\mega\bit\per\second}.
At the same time, Bluetooth is configured as \ac{BLE} beacon, regularly sending \texttt{REQUEST}
signals.

On the logic analyzer installed in the coexistence interface, \textbf{frequent glitches in the \texttt{GRANT} signal can be
observed while Wi-Fi is under high load}, as shown in \autoref{fig:grant-glitch}.
They only occur under traffic and if the coexistence priority mode
is set to \path{WLAN_HIGH} or \path{WLAN_MAXIMIZED}.
While this looks like a power supply issue at first sight, setting the logic
analyzer output to analog mode reveals that the power level drops to \SI{0}{\volt}.
Thus, this is likely a programming error.

\begin{figure}[!h]
	\begin{tikzpicture}[every node/.style={font=\footnotesize}]
		\begin{axis}[
			height=4.5cm, 
			width=\columnwidth, 
			xlabel={Time in \si{\milli\second}},
			ymajorgrids=true,
			xmin=0,
			xmax=1.5,			
			ytick={0, 2, 4},
			yticklabels={\texttt{GRANT}, \texttt{PRIORITY}, \texttt{REQUEST}},
            yticklabel style = {yshift=0.25cm}
			]
			\addplot [const plot, no marks, thin, color=darkblue] table [x=Time, y=REQ, col sep=comma] {data/grant-glitch.csv};
			\addplot [const plot, no marks, thin, color=darkblue!50!black] table [x=Time, y=PRIO, col sep=comma] {data/grant-glitch.csv};
			\addplot [const plot, no marks, thin, color=darkgreen] table [x=Time, y=GRANT, col sep=comma] {data/grant-glitch.csv};
			/legend{REQ,PRIO,GRANT}
		\end{axis}
     	\node[darkgreen] at (5, 1) {\scriptsize{Glitches}};
	\end{tikzpicture}
\caption{\texttt{GRANT} glitches for \texttt{WLAN\_HIGH}/\texttt{MAXIMIZED}.}
\label{fig:grant-glitch}
\end{figure}

Such bugs are surprising given that \emph{Silicon Labs} should have made the same observations with their coexistence development kit.
Optimizing time variation and glitches should
significantly improve their Bluetooth and Wi-Fi performance.

\section{Mitigation}
\label{sec:mitigation}

In the following, we address how coexistence security should be improved and compare this to what has been already patched by manufacturers.
Moreover, we discuss how users can minimize their personal risk of wireless exploitation.

\subsection{Improving Coexistence Security}

Addressing coexistence issues depends on the affected component (hardware or firmware) and how
to control it (e.g., reconfigurable hardware interfaces).
Possible mitigations are:
\begin{itemize}
\item Fixing chip architectures with shared memory components (hardware),
\item stripping metadata and unnecessary information from coexistence protocols (hardware/firmware), and
\item adding plausibility checks on resource claims (firmware).
\end{itemize}


\subsubsection*{Addressing Hardware Issues}

Since our threat model assumes code execution on one wireless core, firmware
patches are only effective when they prevents another core from escalating privileges into the
patched core.
However, the untrusted core could re-enable access to shared hardware components that
are temporarily disabled by firmware configurations. Thus, approaches like firmware
debloating have limited effect~\cite{lightblue}. 
Ideally, coexistence flaws are patched in hardware.

Fixing chip architectures and hardware will take a lot of time until it reaches customers.
Rolling out hardware patches requires a new generation of chips, and as of November 2021,
we have not seen a new \emph{Broadcom} or \emph{Cypress} chip generation that addresses
the shared memory issue. The chips in the latest \emph{iPhone 12} and \emph{13} have a firmware compile
date of October 2019, which is prior to our report of the shared memory code execution flaw.
Thus, we expect hardware-based patches not earlier than the \emph{iPhone 14} release.

%


\subsubsection*{Changing Coexistence Protocols}

Changes to protocol implementations should not reduce the performance
under normal operation. However, additional plausibility checks and countermeasures on
suspected attack attempts and generally stripping information from these protocols
pose a vast \textbf{potential to impact performance negatively}.

Moreover, protocol changes require that all wireless chips update protocol implementations at once
to maintain inter-chip compatibility.
Even if they might be patchable in firmware, parts of these protocols might be implemented in
hardware, i.e., firmware can only access abstract packet type and time information but
is not able to manipulate raw coexistence protocol signals or adjust jitter to packet time slots.

\subsection{Vendor Patches and Timeline}
\label{ssec:patches}

The complexity of coexistence patches, including the parts that are theoretically patchable as
they are implemented in firmware and not hardware, raises an important question: How did
vendors apply patches?

\subsubsection*{Responsible Disclosure Timeline}

We reported the first coexistence \ac{DoS} in August 2019. \emph{Broadcom} replied that
they would add protected register access to prevent coexistence reconfiguration via Bluetooth.
However, ARM Cortex M3/M4 these chips are based on does not have such a feature.
In January 2020, the issue was still unpatched on an up-to-date \emph{iOS}.
Thus, we added \emph{Apple} to the loop, one of the largest customers of \emph{Broadcom} wireless chips,
showing that the bug was not fixed, potentially unpatchable, and according to first measurements, even information
leakage was plausible. Once we had working \acp{PoC} and descriptions for information leakage and code execution,
we started the next round of responsible disclosure in March 2020, which also included \emph{Cypress} and
further selected customers like \emph{Google} and \emph{Samsung}. Moreover, we informed other chip
manufacturers that seemed to have similar issues according to data sheets about the more general nature of
coexistence issues.
Later on, we built \acp{PoC} for \emph{Silicon Labs} chips, which we reported in November 2020 and 
included the Bluetooth SIG.

\subsubsection*{Broadcom Patches}

As of November 2021, more than two years after reporting the first coexistence bug, coexistence attacks, including code execution, still work on up-to-date \emph{Broadcom} chips. This
highlights how hard these issues are to fix in practice. In the following, we outline the usual patch timelines and patches.

\emph{Broadcom} prioritizes customers by the number of affected chips,
with mobile devices patched after 2--4 months. The usual order
is \emph{iOS} with irregular patch releases, then \emph{Samsung}-flavored \emph{Android} with a monthly patch cycle,
followed by \emph{macOS}. To the best of our knowledge, patches need to be requested by \emph{Broadcom's} customers.
This seems to cause a very slow patch timeline for \emph{Linux}-based devices.
Thus, \emph{iPhones} and the newest \emph{Samsung Galaxy S} series are the preferred
devices to check for patches---and have dedicated security teams to contact in case
expected patches are missing.

\subsubsection*{Cypress Patches}

The \emph{Cypress} patch process is independent of \emph{Broadcom} patches.
\emph{Cypress} acquired parts of \emph{Broadcom's} wireless division~\cite{cypressbroadcom}, more precisely the IoT devices.
After the acquisition, code was developed independently. \emph{Cypress} is not necessarily informed by \emph{Broadcom} if there
are vulnerabilities, despite still sharing large parts of the codebase, and patches might be developed independently.
To the best of our knowledge, \emph{Broadcom} never released any publicly visible patch release notes---but \emph{Cypress} released some
in June 2020 and updated the status in October as follows~\cite{cypressspectra}:

\begin{itemize}
\item They claim that the shared RAM feature causing code execution has only been \emph{``enabled by development tools for testing mobile phone platforms''}. They plan to remove stack support for this in the future.
\item The keystroke information leakage is remarked as solved without a patch because \emph{``keyboard packets
can be identified through other means''}.
\item \ac{DoS} resistance is not yet resolved but is in development. For this, \emph{``Cypress plans to implement a monitor feature in the Wi-Fi and Bluetooth stacks to enable a system response to abnormal traffic patterns''}.
\end{itemize}

Dividing chips into mobile devices and IoT devices is misleading, since the actual distinction is
single combo chip versus separate \ac{SECI}-connected chips. The wireless chip in the \emph{Raspberry Pi 3B+/4} series has a shared
RAM and is part of the chips \emph{Cypress} acquired.

\subsubsection*{Escalation Risk Reduction}

Modern operating systems separate the Bluetooth from the Wi-Fi daemon via sandboxing.
Thus, if the Bluetooth daemon is able writing into the Wi-Fi chip RAM, this can already be
considered a threat. This is fixed by preventing the Bluetooth daemon to write into the Bluetooth chip's RAM
after loading firmware patches during driver initialization, which prevents the exploit chain
\emph{Bluetooth daemon$\rightarrow$Bluetooth chip$\rightarrow$\mbox{Wi-Fi} chip} as well as a potential
follow-up escalation into the Wi-Fi daemon. This protection has been put into place since \emph{iOS 13.5},
\emph{Android 10} since the March 2020 release, and \emph{macOS Big Sur}.

Yet, our primary threat defined in \autoref{sec:threatmodel}, which are \textbf{over-the-air attacks via the Bluetooth chip, is not mitigated by current patches}.
Only the interface \emph{Bluetooth daemon$\rightarrow$Bluetooth chip} is hardened, not the shared RAM
interface that enables \emph{Bluetooth chip$\rightarrow$\mbox{Wi-Fi} chip} code execution.
It is important to note that the \textbf{daemon$\rightarrow$chip interface was never
designed to be secure} against attacks. For example, the initial patch could be bypassed with a UART interface
overflow (\emph{CVE-2021-22492}) in the chip's firmware until a recent patch, which was at least applied by \emph{Samsung} in January 2021.
Moreover, while writing to the Bluetooth RAM via this interface has been disabled on \emph{iOS} devices,
the \emph{iPhone 7} on \emph{iOS 14.3} would still allow another command to execute arbitrary addresses in RAM.

Details on how we removed the Bluetooth RAM write protection in order to check if the Bluetooth chip has any other
mitigations in place are provided in Appendix \ref{ssec:appendixwriteram}.
This protection also prevents vendors from using tools like \emph{InternalBlue} to check patches
provided by \emph{Broadcom}. Overall, this mitigation prevents security research while marginally improving security.

%
%
%

\subsection{Personal Risk Minimization}

While hardware-related issues remain unpatched, there are simple measures that \textbf{significantly
reduce the risk of wireless attacks} every user can take:
\begin{itemize}
\item Delete unnecessary Bluetooth device pairings, 
\item remove unused Wi-Fi networks from the settings, and
\item use cellular instead of Wi-Fi at public spaces.
\end{itemize}

An attacker within over-the-air proximity needs to gain code execution on one wireless component initially.
This means that they need to exchange malicious data packets that corrupt memory.
The fewer data exchange possible, the lower the risk of attacks.

Paired Bluetooth devices have special permissions, such as keyboard
input capabilities.
Bluetooth is known for issues in the encryption scheme---in the past three years,
five critical bugs were published by the Bluetooth SIG~\cite{btsigsecurity}.
Yet, even after encryption schemes are broken, an initial
pairing still requires user interaction. When an attacker gets code execution
on a Bluetooth chip, they can get the capabilities of already paired devices---but
cannot add new devices. Hence, it is also recommended to delete unnecessary Bluetooth
pairings. This reduces the risk that an attacker gains capabilities like keyboard input
to use these for further escalation.

Due to the current pandemic situation, many users have Bluetooth enabled
for privacy-preserving contact tracing. The \ac{GAEN} API reduces the risk of exploitation
by using short, fixed-length packets, which are sent without feedback about reception~\cite{gaen}.
Typical memory corruption bugs rely on overflows due to missing length checks and
need a feedback channel to bypass security mechanisms like \ac{ASLR}~\cite{frankenstein}. Thus, we
consider exposure notifications to be reasonably secure.

Smartphones permanently scan for Wi-Fi networks in the background and try
to join them if the network name matches. The most common encryption scheme,
WPA2, only verifies that the client has the correct password~\cite{wpa2}. If an attacker
knows a single valid network name and password configuration of a device,
they can spawn a new access point and get active connections. Even worse,
networks with captive portals do not require any initial password at all.
Thus, users
should remove unused Wi-Fi networks to reduce the risk of data exchange that could
lead to Wi-Fi firmware exploitation.

Some services run in the background even while Wi-Fi and Bluetooth are
disconnected. One example of this is the \ac{AWDL} protocol on \emph{iOS} and \emph{macOS}
devices, which is used to seamlessly transfer files using \emph{AirDrop} and similar
features~\cite{milan}. The initial device scan uses Bluetooth, and the data transfer
takes place via Wi-Fi. Since a lot of \ac{AWDL} functionality is based in the kernel, 
it poses an interesting attack surface, and an exploit for it was published recently~\cite{2020:googleprojectzero}.
Disabling Wi-Fi via the settings menu on \emph{iOS} disables \ac{AWDL}.
This is also a good measure to minimize the risk of joining malicious Wi-Fi networks.

Cellular data plans got more affordable during recent years and cellular network coverage increased.
Disabling \mbox{Wi-Fi} by default and only enabling it when using trusted networks can be considered a good security practice, even if cumbersome. 

\balance

\section{Related Cross-Technology Attacks \& Side Channels}
\label{sec:relwork}

Shared resources and performance optimizations are well-known to introduce side channels
on processors and memory~\cite{rowhammer, spectre, meltdown}.

\emph{Screaming Channels} exploit electromagnetic side channels on mixed-signal chips~\cite{screamingchannels}.
A mixed-signal chip is different from our definition of a combo chip---it just has one digital processing
core, such as an ARM Cortex M4. More precisely, the authors attack a \emph{Nordic Semiconductor} chip with an analog Bluetooth radio frontend.
The digital processing core is running firmware that calculates AES-128 in software, thereby causing
an electromagnetic field. This could already be attacked with classic side-channel attacks in very
close proximity. However, the electromagnetic field couples into the digital radio frontend and
is amplified along with the intended Bluetooth signal, thereby creating a \emph{Screaming Channel} 
that leaks similar information over a distance of up to \SI{10}{\meter}.
Compared to our coexistence attacks, observing \emph{Screaming Channels} requires \ac{SDR}-based measurement
equipment, closer and permanent physical proximity. It can only lead to information leakage instead of code execution.

\emph{BLURtooth} exploits the fact that \ac{BLE} and Classic Bluetooth, two modes of operation in Bluetooth
with different lower-layer protocols, support cross-transport key derivation~\cite{antonioli2020blurtooth}.
Since \ac{BLE} and Classic Bluetooth run on the same chip, there is no unauthorized data extraction across
chip boundaries compared to coexistence attacks.

Moreover, it is possible to add cross-technology capabilities to a chip.
For example, \emph{WazaBee} repurposes a Bluetooth radio frontend by patching its firmware to support Zigbee transmissions~\cite{cayre2021wazabee}.
This is possible because \ac{BLE} and Zigbee are very similar on the physical layer.
While such modifications technically enable running two technologies on the same chip, no chip boundaries are violated.


\section{Conclusion}
\label{sec:conclusion}

This paper shows that wireless coexistence comes with a huge attack surface and opens up
various novel attack vectors, which even enable code execution across chips.
While the code execution vulnerability
is rooted in architectural issues of specific chips and uncovering required reverse-engineering efforts,
\ac{DoS} and
information disclosure attacks of a more general nature can directly be
derived from the openly available coexistence specifications.

Wireless coexistence enables new escalation strategies based on
hardwired inter-chip components. Since the attack vector lies directly
between the chips, it bypasses the main operating system.
A full fix will require chip redesigns---current firmware fixes are incomplete.

\section*{Acknowledgment}

We thank \emph{Apple}, \emph{Broadcom}, \emph{Bluetooth SIG}, \emph{Cypress}, \emph{Google}, \emph{Media\-Tek}, \emph{NXP}, \emph{Qualcomm}, \emph{Samsung}, \emph{Silicon Labs}, and \emph{Texas Instruments} for handling the responsible disclosure requests, and \emph{Tesla} for forwarding our information to \emph{Marvell}.

\iftoggle{blinded}{Further acknowledgments are blinded for review.} 
{We thank Dennis Heinze for porting \emph{InternalBlue} to \emph{iOS} and testing a subset of the CVEs, and Dennis Mantz for testing, and Jan Ruge for the \emph{Frankenstein} support.
We thank Matthias Schulz for adding support for the \emph{Samsung Galaxy S10e} to \emph{Nexmon}.
Moreover, we thank Marco Cominelli for the assistance in 
capturing a trace and Matthias Gazzari for the keyboard expertise. We thank Dominik Maier, Max Maass, Arash Asadi, and Luis Henrique de Oliveira Alves for proofreading and
Ralf-Philipp Weinmann for the feedback.
We thank the anonymous shepherd for the very constructive dialog.

This work has been supported by BMBF Open6GHub, DFG SFB 1053 MAKI, HMWK LOEWE emergenCITY, and BMBF/HMWK National Research Center for Applied Cybersecurity ATHENE.}

\section*{Availability}
Our \acp{PoC} are based on \emph{InternalBlue} scripts and \emph{Nexmon} patches.
They are publicly available as part of the \emph{InternalBlue} project on
\url{https://github.com/seemoo-lab/internalblue}.

\newpage



%
%
%

\bibliographystyle{IEEEtranS} 
\bibliography{bibliographies}

\clearpage 

\newpage

\def\thesection{\Alph{section}}
\setcounter{section}{0}
\section{Appendix}

In this appendix, we provide additional information about the \emph{Broadcom} and \emph{Cypress} vulnerabilities
presented in this paper.
The details in this appendix include vendor-specific address ranges to leak information or
execute code (see \autoref{ssec:appendixmapping}) as well as \ac{SECI} protocol details (see \autoref{ssec:appendixseci}).
Finally, we provide information about how we analyzed and removed the Bluetooth write RAM mitigation on various firmware (see \autoref{ssec:appendixwriteram}).

\subsection{WLAN RAM Sharing}
\label{ssec:appendixmapping}

The so-called \emph{``WLAN RAM Sharing''}, previously discussed in \autoref{ssec:archrev}, enables code execution.
The address range of this mapping is shown in \autoref{tab:wlmap}.
Wi-Fi RAM is mapped into Bluetooth starting at \path{wlan_mem_base}.


\subsubsection*{Reading RAM}
In the reading direction, the full memory region is not mapped all the time.
When just reading in the memory area starting at \path{wlan_mem_base},
it tends to show some repeating chunks and a lot of zero padding.
If a block is currently not ready but read by Bluetooth, this sometimes crashes the Bluetooth firmware.
The exact behavior of this memory region depends on the chip. While we are
able to read valid Wi-Fi memory chunks on some devices, reading access is not always stable.

\subsubsection*{Writing RAM}

The underlying hardware abstracts writing to shared RAM much more transparently.
Without having to deal with any special control registers, it is possible to write to
the memory-mapped area via Bluetooth and the written values will immediately appear in the Wi-Fi RAM.

When writing to the Bluetooth chip addresses starting from \texttt{0x680000},
these values appear in the Wi-Fi memory dump starting from \mbox{\path{0x10000}}, the \path{wlan_mem} area.
The debug memory dump is relative to the Wi-Fi chip's RAM address, it is not an absolute address.
The memory dump has an offset of \texttt{0x170000}, meaning that Bluetooth
can control the Wi-Fi area, mapped at \texttt{0x180000}.

\begin{table}[!b]
\renewcommand{\arraystretch}{1.3}
\caption{Shared RAM mapping.}
\label{tab:wlmap}
\centering
\scriptsize
\begin{tabular}{|l|l|l|l|}
\hline
\textbf{Bluetooth Symbol} & \textbf{Bluetooth} & \textbf{Wi-Fi} & \textbf{Wi-Fi Region} \\
\hline
\path{wlan_mem_base} & \texttt{0x680000} & \texttt{0x180000} & \path{shared_base} \\ 
\path{wlan_mem} area & \texttt{0x680000}--... & \texttt{0x180000}--... & \path{wlan_mem} area \\ 
\hline
\end{tabular}
\end{table}

%
%
%

\subsubsection*{Code Execution Examples}
\label{ssec:appendixexec}

On the \emph{BCM4377B3} chip of a \emph{MacBook 2020} model on \emph{macOS 10.15.7}, the latest version of \emph{Catalina},
code written to the Bluetooth address \texttt{0x68cbfc} is executed by Wi-Fi. 
The Wi-Fi firmware running on a \emph{BCM43475B1} chip with the \emph{Android 10}
patch level of March 2020 
is exploitable by writing an arbitrary value to the address \texttt{0x681024} via Bluetooth.
These are firmware version dependent examples, and generally, a crash caused by writing to this region
indicates exploitability on a chip.

\subsubsection*{Kernel Panics}

Writing random bytes to the shared memory on the \emph{BCM43475B1} chip in a \emph{Samsung Galaxy S10/S20} causes various crashes,
including a kernel panic due to the locked PCIe Wi-Fi communication. 
We can produce kernel panics on \emph{Android 9} with a patch level of May 2019. On a more recent March 2020 \emph{Android 10} release, less severe issues
within the driver occur.
The phone is no longer able to transmit packets without a manual reboot. At least, the kernel does not panic and the device does not reboot unintended.

The slightly older \emph{BCM4347B0} chip in \emph{Samsung Galaxy S8} devices on \emph{Android 9}
reacts similarly including kernel panics.
The kernel panics are due to multiple failed attempts of trying to power up \mbox{Wi-Fi}.
However, most of the time, only the Wi-Fi core crashes, and sometimes also \path{wpa_supplicant}, which manages Wi-Fi connections on \emph{Android}.
Note that the \emph{Samsung Galaxy S8} still receives quarterly security updates but is not supported beyond \mbox{\emph{Android 9}}. 

On \emph{iOS}, the majority of kernel panics occurs due to Wi-Fi driver hangs.
The most unstable Wi-Fi implementation is on the \emph{iPhone 6}, which
panics immediately on \emph{iOS 12.5.1} when writing random bytes to the shared memory.
On an \emph{iPhone 8} on \emph{iOS 13}, we could create malformed PCIe \ac{IOMMU} requests.
Hardware components like the \ac{IOMMU} are model-specific, and hangs in the kernel can be fixed in software.
We could not reproduce kernel panics on the \emph{iPhone 11} on \emph{iOS 14}.

\subsection{SECI Protocol Internals}
\label{ssec:appendixseci}

In the following, we detail \ac{SECI} protocol reverse-engineering results required for
reproducing our findings. More specifically, we detail how \ac{SECI} is mapped and accessed 
within the Bluetooth and Wi-Fi firmware.

\begin{table}[!b]
\renewcommand{\arraystretch}{1.3}
\caption{Coexistence register mapping in Bluetooth.}
\label{tab:btmapping}
\centering
\scriptsize
\begin{tabular}{|l|l|l|}
\hline
\textbf{Address} & \textbf{Name} & \textbf{Function}\\
\hline
\texttt{0x650000}--\texttt{0x6507ff} & -- & GCI region. \\
\texttt{0x650060} & \path{gci_input}  & Value received from Wi-Fi. \\
\texttt{0x650160} & \path{gci_output}  & Value sent to Wi-Fi. \\
\hline
\end{tabular}
\end{table}

\subsubsection*{Bluetooth Hardware Mapping}
\label{sssec:secibtmapping}

Symbols belonging to the \emph{CYW20719} Bluetooth module can be extracted from \emph{WICED Studio 6.2}~\cite{wiced}. 
The most important mappings extracted from these symbols are listed in \autoref{tab:btmapping} and explained in the following.

For each packet sent by Wi-Fi, Bluetooth receives a \SI{64}{\bit} value in \path{gci_input}. In the opposite direction, Bluetooth sends information to Wi-Fi via \path{gci_output}.
The values written to \path{gci_output} and received via \path{gci_input} similar to those that are visible
when intercepting the protocol with a logic analyzer. Thus, instead of
wiretapping the \ac{SECI} physical layer, an attacker can also intercept the coexistence registers and
obtain the same information.

\subsubsection*{Wi-Fi Hardware Mapping}
\label{sssec:seciwifimapping}
\emph{Broadcom} Wi-Fi firmware is well-docu\-mented within the \emph{Nexmon} binary patching framework,
which is specifically designed for those chips~\cite{schulz2018}.
Moreover, source code releases for some devices exist, such as the one by \emph{Asus}
for the \emph{RT-AC86U} router~\cite{asusleak}. Based on this knowledge, we  
analyze the Wi-Fi coexistence implementation.

The main Wi-Fi firmware running on an ARM core
maps the \ac{SECI} registers in the same order as in Bluetooth 
in a struct called \path{chipcregs_t}.
The firmware uses the macro \path{NOTIFY_BT_CHL} to notify Wi-Fi about its current \SI{2.4}{\giga\hertz} channel and applies the channel bandwidth with the macro \path{NOTIFY_BT_BW_20}.
These values are combined into one byte, with the first half representing Wi-Fi channels \mbox{1--11} or 0 for no channel, and the second half being 2 or 4 for \SI{20}{\mega\hertz}
or \SI{40}{\mega\hertz}.
Writing these values then triggers the process of channel blocklisting in the Bluetooth
firmware. 

In addition to the firmware running on the ARM core,
\mbox{Wi-Fi} has a high-performance \emph{D11} core.
This proprietary core uses a special assembly language. It implements a state machine parsing all timing-critical low-level information.
The \emph{D11} core maps the Bluetooth \path{gci_output} register as input
to \texttt{0xaeb} and \texttt{0xaec}. It polls them every two Bluetooth clock cycles, with one cycle being defined as \SI{0.625}{\milli\second}~\cite[p.~2318]{bt52}.
Thus, this is the maximum time resolution we can observe via Wi-Fi.

%

\subsection{Write RAM Mitigation}
\label{ssec:appendixwriteram}

The base firmware in the ROM is temporarily patched in the RAM by the operating system's driver
during chip initialization. The operating system can issue a \path{Write_RAM} \ac{HCI} command,
which is required to install patches. The \emph{InternalBlue} experimentation framework features
the same functionality~\cite{mantz2019internalblue}. Using \emph{InternalBlue}, we can
directly write to the Bluetooth RAM that jumps to the \mbox{Wi-Fi} RAM. We require this test setup---we responsibly disclose all our bugs and, thus, do not have a working over-the-air exploit at the
time of testing.

\balance

After releasing our escalation path Bluetooth daemon$\rightarrow$Bluetooth\,chip$\rightarrow$Wi-Fi,
\emph{Broadcom} disabled the \path{Write_RAM} after firmware initialization. We can remove this patch to
confirm that the actual privilege escalation remains unpatched. \ac{HCI} commands are handled by the
function \path{bthci_cmd_GetDefaultCommandHandler}. Since ROM patches are limited and all vendors have individual \ac{HCI} patches for their proprietary features,
the \ac{HCI} command handler first checks for a variable function table in RAM. Binary diffing can automatically
locate the \ac{HCI} command handler. The vendor-specific command handlers differ a lot, however,
they typically use similar assembler instructions for skipping the \path{Write_RAM} command. These can
be replaced by a non-existent handler.

On \emph{iOS}, we first noticed \path{Write_RAM} being disabled on \emph{iOS 13.5}, but it was still
present on \emph{iOS 13.3}. In \emph{iOS 13.5}, the patches still ship as separate \texttt{.hcd} files,
which is the common \emph{Broadcom} patch format. Since \emph{iOS 13.6} and also in the current \emph{iOS 14.3},
the \texttt{.hcd} files are embedded into the \path{BlueTool} binary, which applies these patches to the chip. Writing
to Wi-Fi RAM via Bluetooth is still possible on \emph{iOS 14.3} after removing the \path{Write_RAM} patch.

\emph{Samsung} also uses the \texttt{.hcd} format. Starting on the \emph{Android 10} March 2020 and
\emph{Android 9} June 2020 release or slightly earlier, they removed the \path{Write_RAM} and \path{Read_RAM} 
handlers. The \emph{Android 11} January 2021 release for the \emph{Samsung Galaxy S20 5G}
integrates further validations. The \texttt{BRCMcfgD} patch configuration region, which can still be changed, enforces checks
on the patch region. Thus, we manually craft a \texttt{BRCMcfgD}-based patch, which confirms that
the shared RAM region still is present.

\emph{macOS} removed \path{Write_RAM} the slowest. On all \emph{Catalina} versions, as tested on three
different chips, \path{Write_RAM} still works. This handler was removed in the \emph{Big Sur} release.
Patches are stored in a slightly different \texttt{.hex} format on the read-only \path{/System} volume.
Changing files on this volume is possible but requires disabling \ac{SIP}, which should only be
done on testing devices not containing any user data.

For the \emph{Raspberry Pi}, which is the most popular platform on \emph{Linux}, the situation is even worse.
The patch slots for this specific
chip are exhausted. Patches are typically not shipped for \emph{Raspberry Pis}.

The minimal \path{Write_RAM} patch against coexistence escalation only covers operating system to chip attacks.
It slows down security testing and patch confirmation outside of \emph{Broadcom}.
\textbf{Over-the-air attackers can still escalate their privileges from Bluetooth to Wi-Fi.}

\end{document}